\documentclass[aps,prd,preprint,amsmath,amssymb,showpacs,preprintnumbers,superscriptaddress]{revtex4-2}
\usepackage{CJK}
\usepackage{graphicx}% Include figure files
\usepackage{dcolumn}% Align table columns on decimal point
\usepackage{bm}% bold math
\usepackage{color}
\usepackage{hyperref}% add hypertext capabilities
\begin{document}

\title{Neutrinoless double-beta decay in a finite volume from relativistic effective field theory}% Force line breaks with \\
\author{Y. L. Yang}
\affiliation{State Key Laboratory of Nuclear Physics and Technology, School of Physics, Peking University, Beijing 100871, China}

\author{P. W. Zhao}
\email{pwzhao@pku.edu.cn}
\affiliation{State Key Laboratory of Nuclear Physics and Technology, School of Physics, Peking University, Beijing 100871, China}

\begin{abstract}
The neutrinoless double-beta decay process $nn\rightarrow ppee$ within the light Majorana-exchange scenario is studied using the relativistic pionless effective field theory (EFT) in finite-volume cubic boxes with the periodic boundary conditions.
Using the low-energy two-nucleon scattering observables from lattice QCD available at $m_\pi=300$, 450, 510, and 806 MeV, the leading-order $nn\rightarrow ppee$ transition matrix elements are predicted and their volume dependence is investigated.
The predictions for the $nn\rightarrow ppee$ transition matrix elements can be directly compared to the lattice QCD calculations of the $nn\rightarrow ppee$ process at the same pion masses.
In particular for the matrix element at $m_\pi=806$ MeV, the predictions with relativistic pionless EFT are confronted to the recent first lattice QCD evaluation.
Therefore, the present results are expected to play a crucial role in the benchmark between the nuclear EFTs and the upcoming lattice QCD calculations of the $nn\rightarrow pp ee$ process, which would provide a nontrivial test on the predictive power of nuclear EFTs on neutrinoless double-beta decay.
\end{abstract}

\maketitle

%*********************************************************%
%---------------------Introduction------------------------%
%*********************************************************%
\section{Introduction}
% Neutrinoless double beta decay
Neutrinoless double-beta decay ($0\nu\beta\beta$) is a second-order weak process, where a nucleus decays to its neighboring nucleus by turning two neutrons to two protons, emitting two electrons but no corresponding antineutrinos~\cite{Furry1939Phys.Rev.11841193}.
It violates the lepton number conservation of the Standard Model (SM) of particle physics and, if observed, would confirm that neutrinos are of Majorana nature~\cite{Schechter1982Phys.Rev.D29512954,Zeldovich1981JETPLett.141}.
The theoretical calculations of the $0\nu\beta\beta$ decay rate, combined with experimental searches~\cite{Aalseth2018Phys.Rev.Lett.132502,Adams2020Phys.Rev.Lett.122501,Agostini2020Phys.Rev.Lett.252502,
Albert2018Phys.Rev.Lett.072701,Armengaud2021Phys.Rev.Lett.181802,Arnold2017Phys.Rev.Lett.041801,Gando2016Phys.Rev.Lett.082503,Dai2022Phys.Rev.D32012}, will advance our knowledge of beyond-Standard-Model (BSM) mechanisms that may be responsible for this process.
Moreover, the theoretical predictions of the expected decay rate in the given BSM scenarios will benefit the planned experiments on $0\nu\beta\beta$~\cite{Agostini2023Rev.Mod.Phys.025002}.
However, the theoretical calculations so far suffer from considerable uncertainties, which hamper the interpretation of current experimental limits on $0\nu\beta\beta$ and potential future discoveries.

A major source of uncertainty in calculating the $0\nu\beta\beta$ decay rate is the nuclear matrix elements between the initial and final nuclear states.
For the isotopes of experimental interest for $0\nu\beta\beta$ searches, such uncertainties stem from both the approximations in nuclear-structure models used to solve the many-body wave function, as well as the uncertainties in the $0\nu\beta\beta$ decay operator~\cite{Engel2017Rep.Prog.Phys.046301}.
The latter can be singled out by studying the amplitude of $0\nu\beta\beta$ decay in the two-nucleon sector, i.e., $nn\rightarrow ppee$, since the two-body wave function can be accurately solved in this case.
Although the $nn\rightarrow ppee$ transition is not observed in the free space, it occurs as the key subprocess in the $0\nu\beta\beta$ decay in nuclei.

% Importance of EFT.
Nuclear effective field theories (EFTs) play an important role in addressing the uncertainties of nuclear matrix elements.
Depending on the energy regime of interest, the nuclear EFTs include the chiral EFT~\cite{Weinberg1990Phys.Lett.B288,Weinberg1991Nucl.Phys.B318}, designed for momenta of the order of the pion mass $m_\pi$, and the pionless EFT~\cite{Kaplan1998Physi.Lett.B390,Kaplan1998Nucl.Phys.B329, vanKolck1999Nucl.Phys.A273,Bedaque1998Phys.Lett.B221, Bedaque1998Phys.Rev.CR641,Chen1999Nucl.Phys.A386}, which focuses on momenta well below $m_\pi$ (For recent reviews, see Refs.~\cite{Epelbaum2009Rev.Mod.Phys.17731825, Machleidt2011Phys.Rep.175, Hammer2020Rev.Mod.Phys.025004}).
Based on a power-counting scheme, they can provide an order-by-order routine to improve the $0\nu\beta\beta$ decay operator, which can then be used as inputs in nuclear-structure calculations.
The $0\nu\beta\beta$ decay operators derived from the nuclear EFTs under various BSM scenarios~\cite{Prezeau2003Phys.Rev.D034016,Cirigliano2017J.HighEnergyPhys.82, Cirigliano2018Phys.Rev.C065501, Cirigliano2018Phys.Rev.Lett.202001, Cirigliano2019Phys.Rev.C055504, Dekens2020J.HighEnergyPhys.97,Cirigliano2021Phys.Rev.Lett.172002, Cirigliano2021J.HighEnergyPhys.289, Yang2024Phys.Lett.B138782},
 as well as their impacts on the nuclear matrix elements~\cite{Menendez2011Phys.Rev.Lett.062501,Pastore2018Phys.Rev.C014606, Wang2018Phys.Rev.C031301, Jokiniemi2021Phys.Lett.B136720, Yao2021Phys.Rev.C014315, Weiss2022Phys.Rev.C065501, Belley2024Phys.Rev.Lett.182502},
 have been extensively studied in the literature.
Here, we focus on the widely considered BSM scenario, the light Majorana-neutrino exchange~\cite{Weinberg1979Phys.Rev.Lett.15661570}.
At leading order (LO), according to the Weinberg power counting of chiral EFT, the only contribution to the $0\nu\beta\beta$ decay operator comes from the long-range neutrino exchange~\cite{Cirigliano2018Phys.Rev.C065501}.
However, in the nonrelativistic heavy-baryon formulation, a renormalization-group analysis showed that a $nn\rightarrow pp ee$ contact term should be promoted to LO to ensure renomalizability~\cite{Cirigliano2018Phys.Rev.Lett.202001, Cirigliano2019Phys.Rev.C055504}.
The low-energy constant (LEC) of this contact term is so far highly uncertain because it can only be fixed by fitting to lepton-number-violating data, currently unavailable.
Subsequently, a synthetic datum of the $nn\rightarrow ppee$ amplitude is proposed by developing a generalized Cottingham formula~\cite{Cirigliano2021Phys.Rev.Lett.172002,Cirigliano2021J.HighEnergyPhys.289}.
Nevertheless, this approach introduces model-dependent inputs for elastic intermediate states and neglects inelastic contributions, and the resulting uncertainties need to be further scrutinized.

% Relativistic EFT
In contrast to the nonrelativistic formulation, it was recently found that the uncertain $nn\rightarrow pp ee$ contact term is not required at LO in the relativistic formulation of chiral EFT~\cite{Yang2024Phys.Lett.B138782}.
This formulation is similar to the so-called modified Weinberg approach~\cite{Epelbaum2012Phys.Lett.B338344} developed for two-nucleon scattering, proven to be useful to improve the renormalizability of two-nucleon scattering phase shifts~\cite{Epelbaum2012Phys.Lett.B338344} and the binding energies of few-body systems~\cite{Epelbaum2017Eur.Phys.J.A98,Yang2022Phys.Lett.B137587}.
Thanks to the milder ultraviolet behavior of the relativistic scattering equation, the $nn\rightarrow ppee$ amplitude from long-range neutrino exchange is ultraviolet finite and, thus, can be properly renormalized without promoting the uncertain contact term to the LO decay operator~\cite{Yang2024Phys.Lett.B138782}.
As a result, the $nn\rightarrow ppee$ amplitudes are predicted in a way that is free of model-dependent inputs beyond the EFT framework and can serve as alternative synthetic data to estimate the LEC of the LO $nn\rightarrow pp ee$ contact term in the nonrelativistic formulation.
The predicted $nn\rightarrow ppee$ amplitudes from the relativistic chiral EFT are slightly larger, by about 10\%-40\% than the amplitudes from the generalized Cottingham formula~\cite{Cirigliano2021Phys.Rev.Lett.172002}, depending on the kinematics.
This discrepancy in the amplitudes will in turn propagate to the contact-term contribution in nuclear structure calculations of realistic $0\nu\beta\beta$ decay.
Therefore, it is crucial to validate the $nn\rightarrow ppee$ amplitudes obtained by the existing approaches.

A direct way to validate the EFT predictions on the $nn\rightarrow ppee$ amplitudes is to perform first-principles lattice QCD (LQCD) calculations that incorporate the dynamics of quarks and gluons~\cite{Briceno2015J.Phys.G023101, Cirigliano2019Eur.Phys.J.A197, Kronfeld2019Eur.Phys.J.A196, Cirigliano2020Prog.Part.Nucl.Phys.103771, Drischler2021Prog.Part.Nucl.Phys.103888, Davoudi2021Phys.Rep.174,Cirigliano2022J.Phys.G120502}.
In fact, LQCD has already demonstrated its reach and capability in constraining the pionic amplitudes for the $0\nu\beta\beta$ processes $\pi^-\pi^-\rightarrow ee$ and $\pi^-\rightarrow \pi^+ee$ within the light Majorana-neutrino exchange scenario~\cite{Feng2019Phys.Rev.Lett.022001,Tuo2019Phys.Rev.D094511,Detmold2020arXiv2004.07404} and the $\pi^-\rightarrow \pi^+ee$ process within a heavy-scale scenario~\cite{Nicholson2018Phys.Rev.Lett.172501}.
The LQCD calculations of the processes involving two nucleons are more complicated.
The LQCD computations for the two-neutrino double beta decay $nn\rightarrow ppee\overline{\nu}_e\overline{\nu}_e$~\cite{Shanahan2017Phys.Rev.Lett.062003,Tiburzi2017Phys.Rev.D054505} and, recently, the $0\nu\beta\beta$ decay $nn\rightarrow pp ee$~\cite{Davoudi2024Phys.Rev.D114514} have been achieved.
However, these calculations are currently tractable only at unphysical heavy pion masses, due to the computational cost.
Therefore, to directly benchmark with the available and upcoming LQCD results of the $nn\rightarrow pp ee$ process, the nuclear EFTs need to be implemented at the heavy pion masses same as those in the LQCD calculations. 

In addition, the LQCD calculations of the matrix elements are implemented on a finite-volume lattice in Euclidean spacetime, which means that there are no asymptotic states.
In contrast, the nuclear EFTs calculate the scattering amplitudes within the infinite volume in Minkowski spacetime.
The matching procedure between the Euclidean finite-volume matrix elements and the infinite-volume scattering amplitude for the $nn\rightarrow pp ee$ process has been developed~\cite{Davoudi2021Phys.Rev.Lett.152003, Davoudi2022Phys.Rev.D094502}.
It builds upon the major developments in recent years in accessing transition amplitudes in hadronic physics from the corresponding finite-volume Euclidean matrix elements~\cite{Lellouch2001Commun.Math.Phys.31,Detmold2004Nucl.Phys.A170,Meyer2011Phys.Rev.Lett.072002, Briceno2013Phys.Rev.D094507, Bernard2012J.HighEnergyPhys.23, Briceno2015Phys.Rev.D034501, Briceno2015Phys.Rev.D074509, Briceno2016Phys.Rev.D013008, Christ2015Phys.Rev.D114510, Briceno2020Phys.Rev.D014509, Feng2021Phys.Rev.D034508} and, in particular, the similar procedure developed for two-neutrino double-beta decay $nn\rightarrow pp ee \overline{\nu}_e\overline{\nu}_e$~\cite{Davoudi2020Phys.Rev.D114521}.
First, the connection between the Euclidean matrix element, accessible in LQCD, and its Minkowski counterpart is constructed.
Then, the nonrelativistic formulation of LO pionless EFT is implemented in a finite volume to derive the Minkowski counterpart of the LQCD matrix element.
The Minkowski matrix element depends on the LEC of the LO $nn\rightarrow pp ee$ contact term in the pionless EFT, whose size is still uncertain.
In Ref.~\cite{Davoudi2022Phys.Rev.D094502}, this LEC is estimated from the generalized Cottingham formula~\cite{Cirigliano2021Phys.Rev.Lett.172002,Cirigliano2021J.HighEnergyPhys.289}.
However, this estimation is only applicable at the physical pion mass.
At the unphysical heavy pion mass tractable in the LQCD calculations, the nonrelativistic pionless EFT needs to fit the LQCD results of the $nn\rightarrow pp ee$ matrix elements to fix the unknown LEC of $nn\rightarrow pp ee$ contact term, instead of making predictions.

In this work, we implement the relativistic formulation of pionless EFT in a finite volume at various pion masses, to predict the Minkowski matrix elements that can be directly compared to the LQCD calculations of the $nn\rightarrow pp ee$ process.
Different from the nonrelativistic case, we estimate the LEC of the LO $nn\rightarrow pp ee$ contact term by integrating out the pion contributions to the long-range neutrino potential in the relativistic chiral EFT.
This is possible because the long-range $nn\rightarrow pp ee$ amplitudes are renormalizable in the relativistic chiral EFT~\cite{Yang2024Phys.Lett.B138782}.
For the physical pion mass, we compare the scattering amplitudes from the relativistic formulation with the previous nonrelativistic results in Ref.~\cite{Davoudi2022Phys.Rev.D094502} with the contact term fixed by the generalized Cottingham formula~\cite{Cirigliano2021Phys.Rev.Lett.172002}.
For the unphysical pion masses $m_\pi=300$, 450, 510, and 806 MeV, using the available two-nucleon observables from the LQCD calculations ~\cite{Yamazaki2012Phys.Rev.D074514,Beane2013Phys.Rev.D034506, Yamazaki2015Phys.Rev.D014501,Illa2021Phys.Rev.D054508,Amarasinghe2023Phys.Rev.D094508} as inputs, 
we present the EFT predictions of the LO Minkowski matrix elements in finite volumes, as well as the scattering amplitude in the infinite volume.
Finally, we compare the EFT predictions and the first LQCD evaluation of matrix element at $m_\pi=806$ MeV~\cite{Davoudi2024Phys.Rev.D114514} with the same finite volume.
The present results can be used in the future benchmarks between the nuclear EFTs and the LQCD calculations of the $nn\rightarrow pp ee$ process, which would provide a solid assessment of the predictive power of nuclear EFTs on $0\nu\beta\beta$ decay.
The results are also expected to be instructive for the analysis of the systematic uncertainties in the future LQCD calculations of the $nn\rightarrow pp ee$ process.

%*********************************************************%
%----------------Theoretical framework--------------------%
%*********************************************************%
\section{Theoretical framework}
In section~\ref{sec.IIA}, we describe the relativistic formulation of pionless EFT~\cite{Epelbaum2012Phys.Lett.B338344,Yang2022Phys.Lett.B137587, Yang2024Phys.Lett.B138782} employed to evaluate the $nn\rightarrow pp ee$ amplitudes.
In section~\ref{sec.IIB}, we briefly introduce the implementation of the relativistic formulation of pionless EFT in a finite volume, closely following the derivations in Refs.~\cite{Davoudi2021Phys.Rev.Lett.152003, Davoudi2022Phys.Rev.D094502} for the nonrelativistic case.

\subsection{Relativistic formulation of pionless EFT}\label{sec.IIA}
The pionless EFT is based on the tenet that the few-nucleon processes at very low energies, i.e., $Q\ll m_\pi$, are not sensitive to the details associated with pions or other meson exchanges~\cite{Hammer2020Rev.Mod.Phys.025004}.
Then, all mesons can be integrated out and the effective Lagrangian contains nucleon and lepton degrees of freedom, organized according to the number of derivatives.
The observables are expanded in $Q/m_\pi$, where $Q$ is the low-energy scale of the order of the binding momentum $\gamma=\sqrt{m_N B_{NN}}$ or of the inverse of the scattering length $a$. 
At LO, the effective Lagrangian reads
\begin{equation}
    \begin{split}
    \mathcal{L}_{\Delta L=0}&=\overline{\Psi}({\rm i}\gamma^\mu\partial_\mu-m_N)\Psi-\sum_\alpha\frac{C_\alpha}{2}(\overline{\Psi}\Gamma_\alpha\Psi)^2+\frac{1}{2}\overline{\Psi}(l_\mu\gamma^\mu+g_Al_\mu\gamma^\mu\gamma_5)\Psi,
  \end{split}
\end{equation}
with $\Psi$ the nucleon field , $m_N$ the nucleon mass, LECs $C_\alpha$, $\alpha=S,V,AV,T$, and $\Gamma_\alpha$ the corresponding Dirac gamma matrices.
The nucleons are coupled to the electroweak current $l_\mu$ via both vector coupling $g_V=1$ and axial coupling $g_A=1.27$.
Here, we neglect the dependence of $g_A$ on the unphysical pion mass.
This dependence can be easily taken into account once its value is provided by the LQCD calculations at the corresponding pion mass since it only provides constant factors on the neutrino potential.
The electroweak current reads $l_\mu=-2\sqrt{2}G_F V_{ud}\tau^+\overline{e}\gamma_\mu v_{eL}+{\rm h.c.}$, with the Fermi constant $G_F$ and the $V_{ud}$ element of the Cabibbo-Kobayashi-Maskawa (CKM) matrix~\cite{Cabibbo1963Phys.Rev.Lett.531533,Kobayashi1973Prog.Theor.Phys.652657}.

After expanding the nucleon field in terms of the free Dirac spinors $u(\bm p,s)$  with positive energies in momentum space and keeping only the leading term of $u(\bm p,s)$ expanding in powers of small momenta $\bm p$, the LO strong potential takes the following form,
\begin{equation}
    \begin{split}
    	V_S(\bm p',\bm p)&=\frac{m_N}{\omega_{p'}}\frac{m_N}{\omega_{p}}[C_S+C_V-(C_{AV}-2C_T)\bm\sigma_1\cdot\bm\sigma_2],
    \end{split}
\end{equation}
where $\bm p'$ and $\bm p$ are the nucleon's final and initial momenta in the center-of-mass frame, respectively, and $\omega_p=(m_N^2+p^2)^{1/2}$.
At the leading order of pionless EFT, the interaction only contributes to the $s$ wave.
Here, we consider the $^1S_0$ channel relevant for the $nn\rightarrow ppee$ process, in which $\bm\sigma_1\cdot\bm\sigma_2=-3$, and the LEC $C=C_S+C_V+3(C_{AV}-2C_T)$ determines the interaction strength in this channel.
As in the nonrelativistic case, the LEC $C$ scales as $\mathcal{O}(4\pi/(m_N Q))$ so that the LO  amplitudes consist of a resummation of the LO interaction $V_S$ and the low-energy pole in the two-nucleon $^1S_0$ amplitude can be reproduced.
In particular, for the two-nucleon scattering process, the amplitude reads 
\begin{equation}\label{eq.LS}
	{\rm i}\mathcal{M}_S(E)=-{\rm i}V_S(p, p)+\int\frac{{\rm d}^3 k}{(2\pi)^3}(-{\rm i})V_S(p, k)\frac{{\rm i}}{E-2\omega_k+{\rm i}0^+}{\rm i}\mathcal{M}_S(E).
\end{equation}
with $p=\sqrt{\frac{1}{4}E^2-m_N^2}$.
After regularizing the contact term with a separable regulator, $V_S(p',p)\rightarrow V_S(p',p)f_\Lambda(p')f_\Lambda(p)$, with momentum cutoff $\Lambda$, the scattering amplitude takes the form
\begin{equation}\label{eq.M_S}
    \mathcal{M}_S(E)=-\frac{1}{C_\Lambda^{-1} - I_\Lambda(E)},
\end{equation}
where the ``bubble integral" reads
\begin{equation}\label{eq.Ilam}
	I_\Lambda(E) = \int\frac{{\rm d}^3 k}{(2\pi)^3}\frac{m_N^2}{\omega_k^2}\frac{1}{E-2\omega_k+{\rm i}0^+}[f_\Lambda(k)]^2.
\end{equation}
The divergence of the ``bubble integral" is absorbed into the cutoff dependence of the LEC $C_\Lambda$ and the resulting scattering amplitude is cutoff-independent as $\Lambda\rightarrow\infty$.
For the two-nucleon bound state, the energy eigenvalue $E$ and the momentum-space wave function $\phi_E(\bm p)$ are obtained by solving the following eigen equation
\begin{equation}\label{eq.eigen}
	(E-2\omega_p)\phi(\bm p)=\int\frac{{\rm d}^3 k}{(2\pi)^3}V_S(p,k)\phi(\bm k).
\end{equation}

Beyond LO, the strong interactions arise from the Lorentz-invariant contact Lagrangian with an increasing number of derivatives (See Ref.~\cite{Xiao2019Phys.Rev.C024004} for the expressions of the contact Lagrangian up to fourth-order derivatives).
Here, we consider the ordering of the contact terms in the relativistic pionless EFT to be the same as its nonrelativistic counterpart, because they both should reproduce the effective range expansion order by order, $p\cot\delta(p)=-\frac{1}{a}+\frac{1}{2}rp^2\ldots$, with the ellipse denoting high-order momentum dependences.
The subleading contribution is given by the effective range $r\sim O(m_\pi^{-1})$, and it is of the order $O(Q/m_\pi)$ relative to the LO contribution.
Once properly renormalized, the difference between the LO relativistic and nonrelativistic pionless EFTs is that the former includes higher-order terms dictated by the Lorentz invariance.
For example, the LO relativistic EFT yields a small effective range of $r\sim O(m_N^{-1})$, while the nonrelativistic theory yields exactly zero effective range $r=0$.

In this work, we consider the standard mechanism of $0\nu\beta\beta$ decay, in which the lepton number violation at low energies is dominated by a Majorana mass term
\begin{equation}
  \mathcal{L}_{\Delta L=2}=-\frac{m_{\beta\beta}}{2}\nu^T_{eL} C \nu_{eL},
\end{equation}
where $C={\rm i}\gamma_2\gamma_0$ denotes the charge conjugation matrix and $m_{\beta\beta}$ the effective neutrino mass.
Following naive dimension analysis, the LO neutrino potential is contributed only by the long-range exchange of a potential neutrino, 
\begin{equation}\label{eq.VnuL}
    V_{\nu{\rm L}}(\boldsymbol p',\boldsymbol p)=(1+3g_A^2)\frac{m_N}{\omega_{p'}}\frac{m_N}{\omega_{p}}\frac{1}{|\boldsymbol p'-\boldsymbol p|^2},
\end{equation}
and its contribution is $O(Q^{-2})$.
Similar to the strong potential, there are $m_N/\omega_p$ factors coming from the expansion of the nucleon field without performing the nonrelativistic reduction.
However, in pionless EFT, it is known that contact operators in weak processes are typically enhanced when $S$ waves are involved~\cite{Bedaque2002Ann.Rev.Nucl.Part.Sci.339,Hammer2020Rev.Mod.Phys.025004}.
This leads to the existence of a contact term in the LO neutrino potential--despite being expected two orders down in the naive dimension analysis
\begin{equation}\label{eq.Vct}
    V_{\nu{\rm CT}}(\boldsymbol p',\boldsymbol p)=-2\frac{m_N}{\omega_{p'}}\frac{m_N}{\omega_{p}}g_\nu^{NN}.
\end{equation}
with $g_\nu^{NN}$ the corresponding LEC.
However, different from the nonrelativistic case, $g_\nu^{NN}$ does not serve as a counterterm that absorbs the ultraviolet divergence, since the LO long-range amplitude itself is ultraviolet finite~\cite{Yang2024Phys.Lett.B138782}.
As a result, the contact-term contribution to the LO amplitude is finite in relativistic pionless EFT and can be estimated by integrating out the pion contributions in the relativistic chiral EFT, as we will demonstrate below.

\begin{figure}[!htbp]
    \centering
    \includegraphics[width=0.8\textwidth]{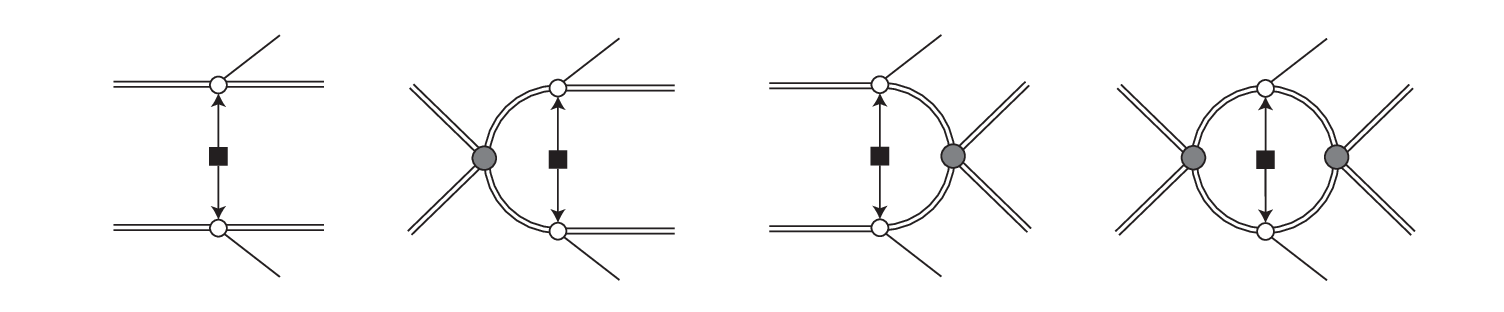}
    \caption{Diagrams contributing to the LO long-range $nn\rightarrow ppee$ amplitude.
    The amplitude $\mathcal{M}^{({\rm int})}_{\rm 0\nu}$ excluding neutrino exchanges on the external legs is shown by the last diagram.
    The double and plain lines denote nucleons and leptons, respectively.
    The squares denote insertions of neutrino potential $V_\nu$.
    The empty circles denote the nucleon axial and vector currents coupled to neutrino exchange.
    The gray circles represent the LO two-nucleon scattering amplitude $\mathcal{M}_S$.}
    \label{fig1}
\end{figure}

The LO long-range $nn\rightarrow ppee$ amplitude
 can be schematically written as 
\begin{equation}\label{eq.M0nu}
    \begin{split}
    \mathcal{M}_{0\nu,{\rm L}}&=-m_{\beta\beta}(1+3g_A^2)\left(V_{\nu {\rm L}}-\mathcal{M}_S I^\infty -\overline{I}^\infty\mathcal{M}_S+\mathcal{M}_S J^\infty\mathcal{M}_S\right),
    \end{split}
\end{equation}
with $\mathcal{M}_S$ is the LO two-nucleon scattering amplitude [Eq.~(\ref{eq.M_S})], and $I^\infty$ and $J^\infty$ are the one- and two-loop integrals with a neutrino exchange.
Here, the four terms correspond to the four diagrams depicted in Fig.~\ref{fig1}.
The first three terms are the contributions in which a neutrino propagates between two external nucleons.
The last term, where a neutrino propagates between two nucleons dressed by strong interactions in both the initial and final states, is the subject of matching to LQCD matrix elements~\cite{Davoudi2021Phys.Rev.Lett.152003},
\begin{equation}\label{eq.MintL}
    \mathcal{M}_{0\nu,{\rm L}}^{({\rm int})}(E_f, E_i)=-m_{\beta\beta}(1+3g_A^2)\mathcal{M}_S(E_f)J^\infty(E_f,E_i)\mathcal{M}_S(E_i),
\end{equation}
where $J^\infty$ takes the form,
\begin{equation}\label{eq.Jinf}
    J^\infty(E_f,E_i)=\int\frac{{\rm d}^3 k_1}{(2\pi)^3}\frac{{\rm d}^3 k_2}{(2\pi)^3}
    \frac{m_N^2}{\omega_{k_1}^2}\frac{1}{E_f-2\omega_{k_1}+{\rm i}0^+}
    \frac{m_N^2}{\omega_{k_2}^2}\frac{1}{E_i-2\omega_{k_2}+{\rm i}0^+}
    \frac{1}{|\bm k_1-\bm k_2|^2}.
\end{equation}
This two-loop integral is ultraviolet finite in the present relativistic formulation, but divergent in the nonrelativistic case.
This is because, in the nonrelativistic approach, the $1/m_N$ expansion is carried out for the integrand making its ultraviolet behavior more singular and resulting in a logarithmic divergence.

Now, we discuss the contact-term contribution to the LO $nn\rightarrow ppee$ amplitude.
It is well known that the coupling of the axial current to pions gives rise to the Goldberg-Treiman relation between the pseudoscalar and axial contribution to the weak form factor.
In the relativistic chiral EFT, this effect is included in the LO long-range neutrino potential, 
\begin{equation}
    \delta V_{\nu \rm L}(\bm p',\bm p)=-g_A^2\frac{m_N}{\omega_{p'}}\frac{m_N}{\omega_{p}}\frac{|\bm p'-\bm p|^2+2m_\pi^2}{\left(|\bm p'-\bm p|^2+m_\pi^2\right)^2},
\end{equation}
while in the relativistic pionless EFT, the contributions from $\delta V_{\nu \rm L}$ are integrated out and will manifest in the form of a contact term.
Therefore, the LO contact term $g_\nu^{NN}$ can be estimated by the contribution of $\delta V_{\nu \rm L}$.
This is achieved by inserting $\delta V_{\nu \rm L}$ into the four diagrams in Fig.~\ref{fig1} and considering $m_\pi$ as a hard scale.
The first tree-level diagram is just $\delta V_{\nu \rm L}\sim O(m_\pi^{-2})$, i.e., next-to-next-to-leading order (NNLO).
The contribution of the second diagram can be written as $-g_A^2 \delta I^\infty_\pi(E_f,E_i)\mathcal{M}_S(E_i)$ with $\delta I^\infty_\pi(E_f,E_i)$ the one loop integral with an insertion of $\delta V_{\nu \rm L}$,
\begin{equation}
    \delta I^\infty(E_f,E_i)=\int\frac{{\rm d}^3 k}{(2\pi)^3}
    \frac{m_N^2}{\omega_{k}^2}\frac{1}{E_f-2\omega_{k}+{\rm i}0^+}
    \frac{|\bm k-\bm p_i|^2+2m_\pi^2}{\left(|\bm k-\bm p_i|^2+m_\pi^2\right)^2}.
\end{equation}
%\begin{equation}
%    \delta I^\infty(E_f,E_i)=-\frac{M}{4\pi}\frac{1}{m_\pi}\left[\frac{x\sqrt{1-x^2}+(3-4x^2){\rm arccos}(x)}{(1-x^2)^{3/2}}-\frac{\pi(3+4x)}{2(1+x)^2}\right]+O\left(\frac{Q^2}{m_\pi^2}, \frac{Q^2}{M^2}\right)
%\end{equation}
%, with $x=m_\pi/M$.
Expanding this integral around the threshold $E_f=E_i=0$, we have $\delta I^\infty(E_f,E_i)=\delta I^\infty(0,0)+O(Q^2/m_\pi^2)$.
Then, $I_\pi^\infty(0,0)$ is just a function of the hard scales $m_\pi$ and $M$, and dimension analysis determines $\delta I^\infty(E_f,E_i)\sim O\left(m_N/(4\pi m_\pi)\right)$.
Since the scaling of $\mathcal{M}_S$ is $O(4\pi/(m_N Q))$, the contribution of second diagram is subleading, $O((m_\pi Q)^{-1})$.
Following the above analysis, the contribution of the third diagram is also subleading, $O((m_\pi Q)^{-1})$, but the contribution of the fourth diagram is instead LO, $O(Q^{-2})$.
The latter takes the form
\begin{equation}\label{eq.dM0nu}
    \delta\mathcal{M}_{0\nu}^{({\rm int})}(E_f, E_i)=m_{\beta\beta}g_A^2\mathcal{M}_S(E_f)\delta J^\infty(E_f, E_i)\mathcal{M}_S(E_i)
\end{equation}
with 
\begin{equation}
    \delta J^\infty(E_f, E_i)=\int\frac{{\rm d}^3 k_1}{(2\pi)^3}\frac{{\rm d}^3 k_2}{(2\pi)^3}
    \frac{m_N^2}{\omega_{k_1}^2}\frac{1}{E_f-2\omega_{k_1}+{\rm i}0^+}
    \frac{m_N^2}{\omega_{k_2}^2}\frac{1}{E_i-2\omega_{k_2}+{\rm i}0^+}
    \frac{|\bm k_1-\bm k_2|^2+2m_\pi^2}{(|\bm k_1-\bm k_2|^2+m_\pi^2)^2}.
\end{equation}
This two-loop integral has a mass dimension of two and thus scales as 
$O\left(m_N^2/(4\pi)^2\right)$.
As a result, $\delta\mathcal{M}_{0\nu}^{({\rm int})}(E_f, E_i)\sim O(Q^{-2})$ and needs to accounted for by a LO contact term.
The contact term contributes to the amplitude as
\begin{equation}\label{eq.M0nuCT}
    \mathcal{M}_{0\nu,{\rm CT}}^{({\rm int})}(E_f, E_i)=2m_{\beta\beta}\tilde{g}_\nu^{NN}\left(\frac{m_N}{4\pi}\right)^2\mathcal{M}_S(E_f)\mathcal{M}_S(E_i)
\end{equation}
after defining the dimensionless LEC $\tilde{g}_\nu^{NN}$,
\begin{equation}
    \tilde{g}_\nu^{NN}=\left(\frac{4\pi}{m_N C}\right)^2g_\nu^{NN}.
\end{equation}
By matching Eqs.~(\ref{eq.dM0nu}) and (\ref{eq.M0nuCT}) at the threshold $E_f=E_i=0$, the dimensionless LEC can be determined,
\begin{equation}\label{eq.gnuNN}
    \tilde{g}_\nu^{NN}=\frac{(4\pi)^2}{2m_N^2}g_A^2\delta J^\infty(0,0).
\end{equation}

In addition to the contribution originating from the coupling of pions, there could be other unknown short-range contributions to the LO LEC $\tilde{g}_\nu^{NN}$ in the relativistic pionless EFT.
By comparing the pionless and chiral EFT amplitudes, the unknown short-range contributions in the pionless EFT corresponds to the $\tilde{g}_\nu^{NN}$ contact-term contribution in the chiral EFT.
It then follows that, based on Weinberg power counting, $\tilde{g}_\nu^{NN}$ contribution is expect to be suppressed by two orders in the chiral expansion, i.e., $O(Q^2/\Lambda_\chi^2)$ with $\Lambda_\chi\sim m_N$ the break down scale of chiral EFT.
Based on this estimation, the uncertainty of the estimation by Eq.~(\ref{eq.gnuNN}) can be considered subleading, since $Q^2/m_N^2\lesssim Q/m_\pi$ when $m_\pi\leq 806$ MeV (see Table \ref{tab1}).

Finally, adding up the long-range and contact-term contributions, the amplitude is given by
\begin{equation}\label{eq.Mint}
    \mathcal{M}_{0\nu}^{({\rm int})}(E_f,E_i)=-m_{\beta\beta}\mathcal{M_S}(E_f)\left[(1+3g_A^2)J^\infty(E_f,E_i)-2\tilde{g}_\nu^{NN}\left(\frac{m_N}{4\pi}\right)^2\right]\mathcal{M_S}(E_i).
\end{equation}

\subsection{Implementation in a finite volume}\label{sec.IIB}
In this work, we implement the relativistic pionless EFT in a finite-volume (FV) cubic box with spatial extent $L$ and the periodic boundary conditions.
Considering the $nn\rightarrow ppee$ process with the kinematics that the two electrons in the final state are at rest, the Euclidean four-point function accessible from LQCD can be analytically continued to the Minkowski spacetime~\cite{Davoudi2021Phys.Rev.Lett.152003},
\begin{equation}\label{eq.TM}
	\mathcal{T}_L^{(M)}(E_f,E_i)=\int{\rm d}z_0\int_L{\rm d}^3 z[\langle E_f, L|T[\mathcal{J}(z_0,\bm z)S_\nu(z_0,\bm z)\mathcal{J}(0)]|E_i,L\rangle]_L,
\end{equation}
where the subscript $L$ on the spatial integral indicates that the integral is performed over the FV cubic box and $|E, L\rangle$ is the normalized FV $s$-wave two-nucleon state with the center-of-mass energy $E$. Here, $\mathcal{J}$ denotes the hadronic part of the weak current, and the neutrino propagator $S_\nu(z_0, \bm z)$ in a finite volume is given by the Fourier transformation
\begin{equation}
	S_\nu(z_0, \bm z)=\frac{1}{L^3}\sum_{\substack{\bm q\in\frac{2\pi}{L}\mathbb{Z}^3\\\bm q\neq\bm 0}}\int\frac{{\rm d }q_0}{2\pi}{\rm e}^{{\rm i}\bm q\cdot\bm z-{\rm i}q_0z_0}\frac{-{\rm i}m_{\beta\beta}}{q_0^2-\bm q^2+{\rm i}0^+},
\end{equation}
ignoring the small nonzero neutrino mass.
Because the space is limited to a box with the periodic boundary conditions, the momentum modes are discretized, only taking the values with $2\pi/L$ times three-dimensional Cartesian vectors with integer components.
The infrared divergence is regulated by removing the zero-momentum mode of neutrinos.

The energy eigenvalues of the two-nucleon states are also discretized in a finite volume.
Their discrete values $E_n$ are directly related to the two-nucleon scattering amplitudes in the infinite volume, by the L\"uscher quantization condition $\mathcal{F}^{-1}(E_n)=0$~\cite{Luescher1986Commun.Math.Phys.153188,Luescher1991Nucl.Phys.B531578}.
For the $^1S_0$ channel considered in this work, it is
\begin{equation}\label{eq.FVF}
	\mathcal{F}^{-1}(E)=\frac{4\pi}{m_N}
	\left(
	  -\frac{1}{\pi L}\mathcal{Z}_{00}\left[1,\left(\frac{p L}{2\pi}\right)^2\right]+{\rm i}p\right)^{-1}+\mathcal{M}_S(E),
\end{equation}
where $\mathcal{Z}_{00}$ is the zeta function defined in Ref.~\cite{Luescher1986Commun.Math.Phys.153188} and $\mathcal{M}_S(E)$ is the  scattering amplitude defined in Eq.~(\ref{eq.M_S}).
%This mapping is valid up to exponentially suppressed corrections governed by the interaction range.
%Note that here we only consider the scattering amplitudes at low energies which are expected to be dominated by the $S$-wave interaction, and the contributions of all higher-order partial waves are ignored.

For the case in which the initial and final states are ``scattering" states, the Minkowski matrix element $\mathcal{T}_L^{(M)}$ is calculated as following~\cite{Davoudi2021Phys.Rev.Lett.152003},
\begin{equation}
	L^6|\mathcal{T}_L^{(M)}(E_f,E_i)|^2=|\mathcal{R}(E_f)||\mathcal{M}^{({\rm int},L)}_{0\nu}(E_f,E_i)|^2|\mathcal{R}(E_i)|,
\end{equation}
where two FV quantities $\mathcal{M}^{({\rm int},L)}_{0\nu}$ and $\mathcal{R}$ are involved.
The former one corresponds to the amplitude (Eq.~\ref{eq.Mint})  in a finite volume~\cite{Davoudi2021Phys.Rev.Lett.152003,Davoudi2022Phys.Rev.D094502},
\begin{equation}\label{eq.MintL}
    \mathcal{M}_{0\nu}^{({\rm int},L)}(E_f,E_i)=-m_{\beta\beta}\mathcal{M_S}(E_f)\left[(1+3g_A^2)J^L(E_f,E_i)-2\tilde{g}_\nu^{NN}\left(\frac{m_N}{4\pi}\right)^2\right]\mathcal{M_S}(E_i).
\end{equation}
The function $J^L$ resembles the function $J^\infty$ in the infinite volume [Eq.~(\ref{eq.Jinf})], with the momentum integrals replaced by the sum of discrete momentum modes,
\begin{equation}\label{eq.JL}
	J^L(E_f,E_i)=
	\frac{1}{L^6}\sum_{\substack{\boldsymbol{k_1},\boldsymbol{k_2}\in\frac{2\pi}{L}\mathbb{Z}^3\\ \boldsymbol{k_1}\neq\boldsymbol{k_2}
	}}
	\frac{m_N^2}{\omega_{k_1}^2}\frac{1}{E_f-2\omega_{k_1}}
    \frac{m_N^2}{\omega_{k_2}^2}\frac{1}{E_i-2\omega_{k_2}}
    \frac{1}{|\bm k_1-\bm k_2|^2}.
\end{equation}
Here, the imaginary part of the propagator is dropped since now the denominator takes nonzero discrete values.
The FV quantity $\mathcal{R}(E)$ is the generalized Lellouch-L\"uscher residue matrix~\cite{Lellouch2001Commun.Math.Phys.31}, 
\begin{equation}\label{eq.R}
    \mathcal{R}(E_n)=\lim_{E\rightarrow E_n}(E-E_n)\mathcal{F}(E)=\left(\left.\frac{\rm d\mathcal{F}^{-1}}{{\rm d} E}\right|_{E=E_n}\right)^{-1},
\end{equation}
which is the residue of the FV function $\mathcal{F}$ (\ref{eq.FVF}) at FV energies $E_n$.

For the case in which the initial and final states are bound states, $E=2M-B$ with $B>0$, the Minkowski matrix element $\mathcal{T}_L^{(M)}$ is calculated by
\begin{equation}\label{eq.TMb}
    \begin{split}
	\mathcal{T}_L^{(M)}(E_f,E_i)&=m_{\beta\beta}\frac{1}{L^6}\sum_{
	\substack{\bm{k_1},\bm{k_2}\in\frac{2\pi}{L}\mathbb{Z}^3\\ \bm{k_1}\neq\bm{k_2}
	}}\phi^*_{E_f,L}(\bm k_1)V_{\nu{\rm L}}(\bm k_1, \bm k_2)\phi_{E_i,L}(\bm k_2)\\
    &-2m_{\beta\beta}\tilde{g}_\nu^{NN}\left(\frac{MB}{4\pi}\right)^2|\phi(\bm 0)|^2.
    \end{split}
\end{equation}
with $\phi_{E,L}(\bm k)$ the normalized momentum-space wave function of the FV state $|E,L\rangle$ in Eq.~(\ref{eq.TM}). 
On the right hand side, the first and the second terms are respectively the expectations of the long-range neutrino potential $V_{\nu\rm L}$ in Eq.~(\ref{eq.VnuL}) and the contact term in Eq.~(\ref{eq.Vct}) regulated with the same separable regulator as the one for the strong interaction.
The wave function $\phi_{E,L}$ is solved by Eq. (\ref{eq.eigen}) with discrete momentum modes.
Although there is no two-nucleon bound state in the $^1S_0$ channel at the physical pion mass, the above matrix element is relevant for the study at the unphysical pion masses.
At the unphysical pion masses, two nucleons might exhibit a $^1S_0$ bound state predicted by the LQCD calculations~\cite{Yamazaki2012Phys.Rev.D074514,Beane2013Phys.Rev.D034506, Yamazaki2015Phys.Rev.D014501,Illa2021Phys.Rev.D054508}.
Note that there is an ongoing discussion on whether such a bound state exists at the unphysical pion masses, as several newer works~\cite{Francis2019Phys.Rev.D074505,Horz2021Phys.Rev.C014003,Amarasinghe2023Phys.Rev.D094508,Detmold2024arXiv} do not identify such a bound state.

%*********************************************************%
%----------------Numerical details------------------------%
%*********************************************************%
\section{Numerical details}
In this work, we consider several box sizes in the range of $L=8$-16 fm at the physical pion mass, and $L=4$-6 fm at the unphysical pion masses in accordance with the existing LQCD calculations of two-nucleon systems~\cite{Yamazaki2012Phys.Rev.D074514,Beane2013Phys.Rev.D034506, Yamazaki2015Phys.Rev.D014501,Illa2021Phys.Rev.D054508}.
We focus on the scattering amplitudes and the FV matrix elements with equal initial and final energies, $E_f=E_i=E$.
This neglects the masses of the two electrons in the final state of the $nn\rightarrow ppee$ process, since they are much smaller than the intervals between the discrete FV energies.
Throughout this work, the effective neutrino mass $m_{\beta\beta}$ is set to 1 MeV.
For the LO strong potential, we use an exponential regulator $f_\Lambda(k)={\rm e}^{-k^4/\Lambda^4}$.

\subsection{Calculations in the infinite volume}
The momentum integrals (\ref{eq.Ilam}), (\ref{eq.Jinf}), (\ref{eq.eigen}) associated with the calculations in the infinite volume are all calculated numerically using Gaussian quadrature.
For the calculations of the scattering amplitudes, the real and imaginary parts of the two-nucleon free propagator are calculated separately,
\begin{equation}\label{eq.sep}
	\frac{1}{E-2\omega_k+{\rm i}0^+}=P\left(\frac{1}{E-2\omega_k}\right)-{\rm i}\pi\delta(E-2\omega_k).
\end{equation}
Here, $P$ denotes principle-value integral and it is eliminated by a standard subtraction technique~\cite{Gloeckle1983}.
The eigen equation (\ref{eq.eigen}) for the bound states is solved by matrix diagonalization on the Gaussian grids. 

However, special care has to be taken for the infrared singularity of the neutrino potential.
For the calculation of the scattering amplitude $\mathcal{M}_{0\nu}^{({\rm int})}$, inserting the separation (\ref{eq.sep}) into the expression of the two-loop integral $J^\infty$ (\ref{eq.Jinf}), we have
\begin{equation}\label{eq.reJ}
    \begin{split}
    {\rm Re}J^\infty(E_f,E_i)&=\int_0^\infty\frac{k_1^2{\rm d} k_1}{2\pi^2}\left[\frac{m_N^2}{\omega_{k_1}^2}P\left(\frac{1}{E_f-2\omega_{k_2}}\right)\int_0^\infty\frac{k_2^2{\rm d} k_2}{2\pi^2}\right.\\
    &\left.\frac{m_N^2}{\omega_{k_2}^2}P\left(\frac{1}{E_i-2\omega_{k_2}}\right)\frac{1}{4k_1k_2}\ln\frac{(k_1+k_2)^2}{(k_1-k_2)^2}\right]
    -\frac{m_N^2}{32\pi^2}\frac{m_N}{\omega_{p_f}}\frac{m_N}{\omega_{p_i}}\ln\frac{p_f+p_i}{|p_f-p_i|}.
    \end{split}
\end{equation}
On the right-hand side, there are logarithmic divergences in the two terms when $E_f=E_i$.
We introduce a subtraction technique making use of the analytic expression of the following two-loop integral,
\begin{equation}\label{eq.reJnr}
    \begin{split}
    	I^\infty_\Lambda&=\int_0^\Lambda\frac{k_1^2{\rm d} k_1}{2\pi^2}
    P\left(\frac{1}{E_f-k_1^2/m_N}\right)
    \int_0^\infty\frac{k_2^2{\rm d} k_2}{2\pi^2}P\left(\frac{1}{E_i-k_2^2/m_N}\right)
    \frac{1}{4k_1k_2}\ln\frac{(k_1+k_2)^2}{(k_1-k_2)^2}\\
    &=\frac{m_N^2}{32\pi^2}\ln\frac{\Lambda^2-p_i^2}{|p_f^2-p_i^2|}.
    \end{split}
\end{equation}
Denoting the integral term in Eq.~(\ref{eq.reJ}) as $I^\infty$, then we have
\begin{equation}
	{\rm Re}J^\infty(E_f,E_i)=\left(I^\infty - \frac{m_N^2}{\omega_{p_f}\omega_{p_i}}I^\infty_\Lambda\right)+\frac{m_N^2}{\omega_{p_f}\omega_{p_i}}\frac{m_N^2}{32\pi^2}\ln\frac{\Lambda^2-p_i^2}{(p_f+p_i^2)}
\end{equation}
Now, the two terms are both infrared convergent when $E_f=E_i$ and we have confirmed the numerical stability using the above expression.

For the calculations of the matrix element $\mathcal{T}_L^{(M)}$ between bound states, the infrared singularity of the neutrino potential is treated with the Lande subtraction~\cite{Kwon1978Phys.Rev.C932,Landau1983Phys.Rev.C2191,LandePC,Ivanov2001Comp.Phys.Commun.317},
\begin{equation}
	\int_0^\infty\frac{{\rm d}k}{2\pi^2}\frac{1}{4kp}\ln\frac{(k+p)^2}{(k-p)^2}k^2f(k)=
	\int_0^\infty\frac{{\rm d}k}{2\pi^2}\frac{1}{4kp}\ln\frac{(k+p)^2}{(k-p)^2}[k^2f(k)-p^2f(p)]+\frac{1}{8}pf(p)
\end{equation}
with $f(p)$ an arbitrary smooth function.

\subsection{Calculations in a finite volume}
For the calculations of the matrix elements $\mathcal{T}_L^{(M)}$ between the ``scattering" states, the FV quantity $J^L$ (\ref{eq.JL}), which involves summation over the discrete three-momenta $\bm k_1,\bm k_2=\bm n2\pi/L$ with $\bm n\in \mathbb{Z}^3$, is calculated using the method of tail-singularity separation (TSS) described in Ref.~\cite{Tan2008Phys.Rev.A013636}.
In this method, the summation is split into two pieces.
One piece contains the singular contributions around $\bm k_1=\bm k_2$, but it is exponentially decaying when $|\bm k_1|,|\bm k_2|\rightarrow\infty$.
The other piece contains a power-law decaying tail at $|\bm k_1|,|\bm k_2|\rightarrow\infty$, but it is sufficiently smooth so that it can be approximated by its integral counterpart.
Based on this method, we calculate $J^L$ as the following
\begin{equation}
\begin{split}
	J^L(E,E)&=\frac{m_N^2}{4(2\pi)^6}\left\{\sum_{\substack{\bm n_1\in\mathbb{Z}^3}}
	\frac{\tilde{m}}{\tilde{m}^2+n_1^2}\frac{1}{\tilde{\omega}_{n_1}-\tilde{\omega}_p}\left[\mathcal{X}_{\rm sum}(\bm n_1,\tilde{p}^2)+{\rm e}^{-\alpha(n_1^2-\tilde{p}^2)}\mathcal{X}_{\rm int}(n_1^2,\tilde{p}^2)\right]\right.\\
	&\left.+4\pi\int_0^\infty n_1^2{\rm d}n_1 \frac{\tilde{m}}{\tilde{m}^2+n_1^2}\frac{1-{\rm e}^{-\alpha(n_1^2-\tilde{p}^2)}}{\tilde{\omega}_{n_1}-\tilde{\omega}_p}\mathcal{X}_{\rm int}(n_1^2,\tilde{p}^2)\right\}+O({\rm e}^{-\pi^2/\alpha}),
\end{split}
\end{equation}
where $\tilde{m}=2\pi m_N/L$, $\tilde{p}=2\pi p/L$, $\tilde{\omega}=(\tilde{m}^2+\tilde{p}^2)^{1/2}$, and
\begin{equation}
    \begin{split}
	\mathcal{X}_{\rm sum}(\bm n_1,\tilde{p}^2)&=\sum_{\substack{\bm n_2\in\mathbb{Z}^3\\
	\bm n_2\neq \bm n_1}}
	\frac{\tilde{m}}{\tilde{m}^2+n_2^2}\frac{1}{\tilde{\omega}_{n_2}-\tilde{\omega}_p}\frac{1}{|\bm n_1-\bm n_2|^2}
	\left[
	1 - (1-{\rm e}^{-\alpha(n_2^2-\tilde{p}^2)})(1-{\rm e}^{-\alpha|\bm n_1-\bm n_2|^2})
	\right]\\
	\mathcal{X}_{\rm int}(n_1^2,\tilde{p^2})&=
	\int{\rm d}^3n_2\frac{\tilde{m}}{\tilde{m}^2+n_2^2}\frac{1-{\rm e}^{-\alpha(n_2^2-\tilde{p}^2)}}{\tilde{\omega}_{n_2}-\tilde{\omega}_p}\frac{1-{\rm e}^{-\alpha|\bm n_1-\bm n_2|^2}}{|\bm n_1-\bm n_2|^2}-\alpha\frac{\tilde{m}}{\tilde{m}^2+n_1^2}\frac{1-{\rm e}^{-\alpha(n_1^2-\tilde{p}^2)}}{\tilde{\omega}_{n_1}-\tilde{\omega}_p}.
    \end{split}
\end{equation}
In the expression of $\mathcal{X}_{\rm int}$, the second term removes the value at the pole $\bm n_2=\bm n_1$ when replacing $\sum_{\bm n_2}\rightarrow\int{\rm d}^3 n_2$.
We use $\alpha=0.01$ and truncate the integer Cartesian coordinates at $|n_x|,|n_y|,|n_z|\leq 32$ in the present calculations.
Under this condition, we evaluate the geometric constants $\mathcal{X}_2$ with a single sum and $\mathcal{R}_{24}$ with double sums, as defined in Eq.~(A1) of Ref.~\cite{Beane2014Phys.Rev.D074511}, by using the TSS method.
We obtain $\mathcal{X}_2=91.18$ and $\mathcal{R}_{24}=170.9$, which agrees with the corresponding results of Ref.~\cite{Beane2014Phys.Rev.D074511} up to four significant figures.

For the calculations of the matrix element $\mathcal{T}_L^{(M)}$ between the bound states, they can be straightforwardly calculated using Eq.~(\ref{eq.TMb}), once the bound-state wave function $\phi_{E,L}$ is solved.
When solving the bound-state wave function, we truncate the integer sum at $|n_x|,|n_y|,|n_z|\leq \Lambda L/(2\pi)$.
The number of momentum modes could reach several thousand, making direct diagonalization intractable.
Therefore, we use the imaginary-time propagation starting from an initial wave function $\phi_{i}$ to solve for the bound-state wave function
\begin{equation}
    \begin{split}
    	\phi_{E,L}=\lim_{N_\tau\rightarrow\infty}({\rm e}^{-H\Delta\tau})^{N_\tau}\phi_{i},
    \end{split}
\end{equation}
where $\Delta\tau$ is a small imaginary-time step, and ${\rm e}^{-H\Delta\tau}$ is expanded up to $O(\Delta \tau^2)$.
This is particularly efficient because of the separable form of the potential,
\begin{equation}
	H\phi(\bm p)=2\omega_p\phi(\bm p)+C_\Lambda f_\Lambda(p^2)\frac{1}{L^3}\sum_{\bm k\in\frac{2\pi}{L}\mathbb{Z}^3} f_\Lambda(k^2)\phi(\bm k),
\end{equation}
so the numerical complexity scales linearly with the number of momentum modes, instead of cubicly when using direct diagonalization.
We take the initial wave function to be $\phi_{i}(\bm p)={\rm e}^{-p^2/m_N^2}$, but using a different form should not affect the final results once the imaginary-time projection converges.

\subsection{Determination of low-energy constants}
\begin{table}[!htbp]
    \centering
    \caption{The nucleon masses $m_N$ and the two-nucleon binding energies $B_{nn}$ in the $^1S_0$ channel at $m_\pi=300$, 450, 510, and 806.
    The LEC $C_\Lambda$ for the LO strong potential, determined at $\Lambda=50$ fm$^{-1}$ using these inputs, are shown in the fourth column.
    The LEC $\tilde{g}_\nu^{NN}$ for the LO $nn\rightarrow ppee$ contact term, determined by Eq.~(\ref{eq.gnuNN}), are shown in the last column.
    The uncertainty of $\tilde{g}_\nu^{NN}$ from the LQCD inputs is smaller than the last digit and thus not shown.
    At the physical pion mass, since no $^1S_0$ bound state exists, the LEC $C_\Lambda$ is instead fixed by the scattering length $a=-23.74$ fm.
    At $m_\pi=806$ MeV, two sets of LQCD inputs are considered (see text for details).
    }
    \label{tab1}
    \begin{tabular}{ccccc}
    \hline\hline
        $m_\pi$  (MeV) & $m_N$  (MeV) & $B_{nn}$ (MeV) & $C_\Lambda$  (fm$^{2}$) & $\tilde{g}_\nu^{NN}$\\\hline
        140 & 938.9   & - & -0.4157 & 1.66\\
        300~\cite{Yamazaki2015Phys.Rev.D014501} & 1055(4) & 8.5$\left(^{+1.7}_{-0.9}\right)$ & -0.3666$\left(^{+14}_{-24}\right)$ & 1.20\\
        450~\cite{Illa2021Phys.Rev.D054508} & 1226(2) & 13.1$\left(^{+3.0}_{-3.1}\right)$ & -0.2875$\left(^{+29}_{-25}\right)$ & 1.03\\
        510~\cite{Yamazaki2012Phys.Rev.D074514} & 1320(3) & 7.4(1.4) & -0.2483$\left(^{+15}_{-14}\right)$ & 1.00\\
        806~\cite{Beane2013Phys.Rev.D034506} & 1634(18) & 15.9(3.8) & -0.1780$\left(^{+19}_{-17}\right)$ & 0.85\\
        806$^*$~\cite{Amarasinghe2023Phys.Rev.D094508} & 1636(18) & $3.3(7)$ & $-0.1585\left(^{+40}_{-31}\right)$ & 0.85\\
    \hline\hline
    \end{tabular}
\end{table}

At the leading order of relativistic pionless EFT, there are two LECs $C_\Lambda$ and $\tilde{g}_\nu^{NN}$ that need to be determined for predicting the scattering amplitudes and matrix elements for the $nn\rightarrow pp ee$ process.
For $C_\Lambda$, it is the strength of the short-range strong potential and can be fixed by one low-energy observable in the $^1S_0$ channel.
The observables used to determine $C_\Lambda$ are shown in Table \ref{tab1}.
At the physical pion mass, there is no $^1S_0$ bound state, so we fix it using the experimental scattering length $a=-23.74$ fm.
At the unphysical pion masses, several LQCD calculations yielded deeply bound $^1S_0$ two-nucleon state~\cite{Yamazaki2015Phys.Rev.D014501,Illa2021Phys.Rev.D054508,Yamazaki2012Phys.Rev.D074514,Beane2013Phys.Rev.D034506}.
However, many other LQCD studies~\cite{Ishii2012Phys.Lett.B437441,Inoue2012Nucl.Phys.A2843,Horz2021Phys.Rev.C014003,Amarasinghe2023Phys.Rev.D094508,Detmold2024arXiv} did not obtain such bound states, raising concerns on whether or not the previous works correctly determined the two-nucleon spectrum.
There are several explanations for this issue and it is still not completely conclusive whether $^1S_0$ two-nucleon state is bound or unbound at unphysical large pion masses~\cite{Tews2022FewBodySystems67}.
Nevertheless, the results from the present work could be easily adjusted to updated LQCD values of two-nucleon binding energies or scattering lengths.
For example, in Table \ref{tab1}, we considered both the older (without asterisk)~\cite{Beane2013Phys.Rev.D034506} and the latest (with asterisk)~\cite{Amarasinghe2023Phys.Rev.D094508} LQCD values of $B_{nn}$ at $m_\pi=806$ MeV.
For $m_\pi=300$, 400, 510, and 806 MeV, the LEC $C_\Lambda$ is fixed using the two-nucleon binding energy $B_{nn}$ in the infinite volume, extrapolated from the FV energies provided by the LQCD calculations.
For $m_\pi=806^*$ MeV, the LQCD result of $B_{nn}$ is only available at a single finite volume $L=4.6$ fm and, thus, the fitting of the LEC $C_\Lambda$ is performed at the same finite volume.

\begin{figure}[!htbp]
    \centering
    \includegraphics[width=0.5\textwidth]{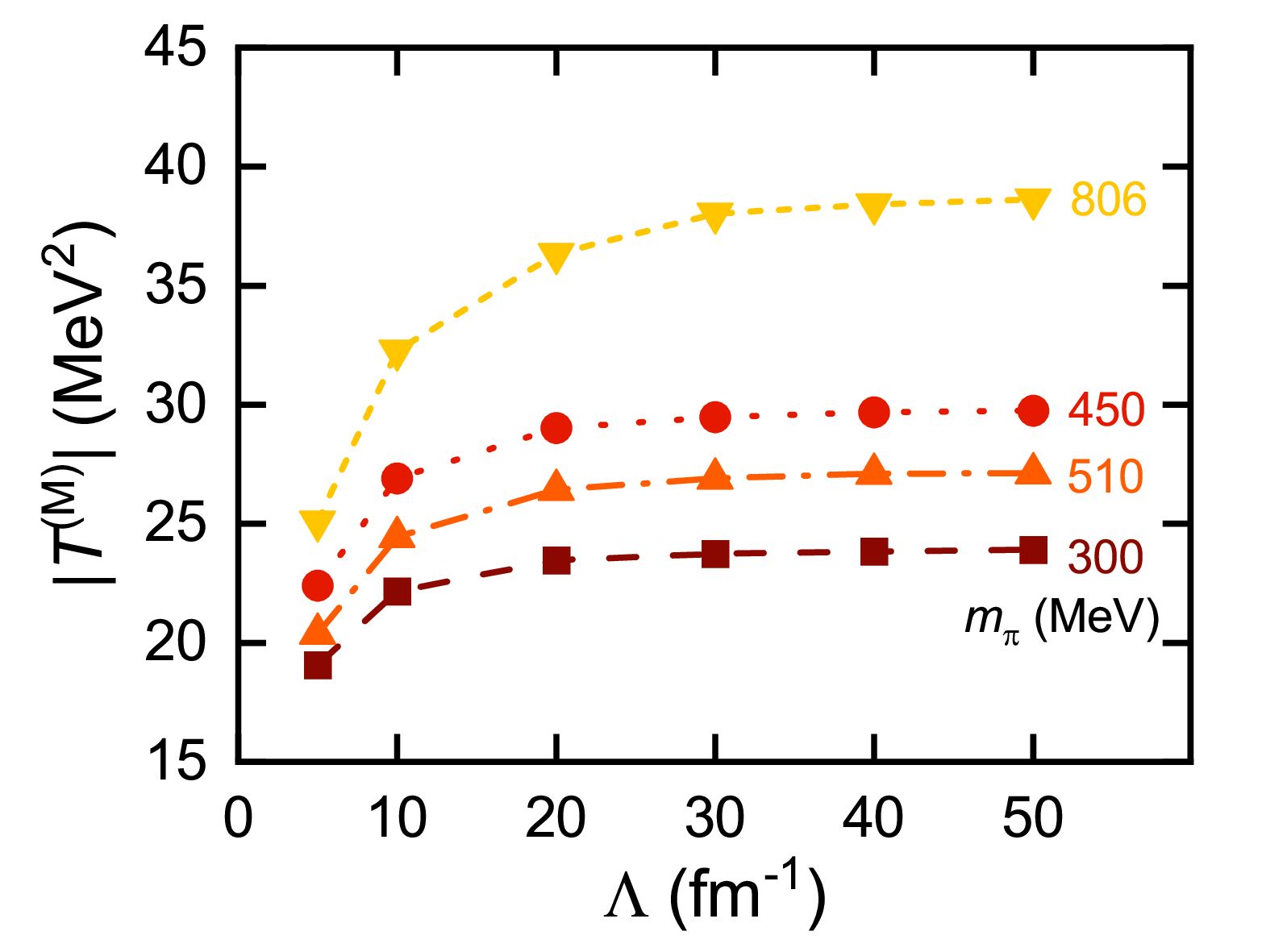}
    \caption{(Color online). The cutoff dependence of the long-range contribution to the LO Minkowski matrix elements $\mathcal{T}^{(M)}$ for the ground-state-to-ground-state transitions in the infinite volume at the pion masses $m_\pi=300$, 450, 510, and 806 MeV, obtained from the relativistic pionless EFT.
    Here, the effective neutrino mass $m_{\beta\beta}$ is set to 1 MeV.}
    \label{fig2}
\end{figure}
In Fig.~\ref{fig2}, we show the cutoff dependence of the long-range contribution to the LO Minkowski matrix elements between the ground states in the infinite volume, $\mathcal{T}^{(M)}=\langle E_0|V_\nu| E_0\rangle$, at the unphysical pion masses.
Here, the LEC $C_\Lambda$ is fitted to the center value of $B_{nn}$ in Table \ref{tab1}.
As expected, the long-range contribution to the matrix elements all converge as the cutoff $\Lambda$ goes to infinity.
For the unphysical pion masses considered here, convergence can be reached at $\Lambda\lesssim50$ fm$^{-1}$ on the $1\%$ level.
As shown in Ref.~\cite{Yang2024Phys.Lett.B138782}, this is also true for the amplitudes at the physical pion mass.
Therefore, we take the amplitudes and matrix elements at $\Lambda=50$ fm$^{-1}$ as the renormalized results in the present study.
The values of LEC $C_\Lambda$ are listed in Table \ref{tab1}.

For the LEC $\tilde{g}_\nu^{NN}$ in the LO contact term in the neutrino potential, it is determined by integrating out the contribution from the coupling of the nucleonic axial current to pions, using Eq.~\ref{eq.gnuNN}.
They are also calculated from the $m_\pi$ and $m_N$ values provided by the experiments or LQCD calulations.
The values of LEC $\tilde{g}_\nu^{NN}$ are listed in Table \ref{tab1}.
Their values are indeed $O(1)$, as expected.
They take positive values and, thus, the contact term contribution reduces the magnitude of the $nn\rightarrow ppee$ amplitude.

%*********************************************************%
%----------------Results and discussion-------------------%
%*********************************************************%
\section{Results and discussion}
\begin{figure}[!htbp]
    \centering
    \includegraphics[width=0.6\textwidth]{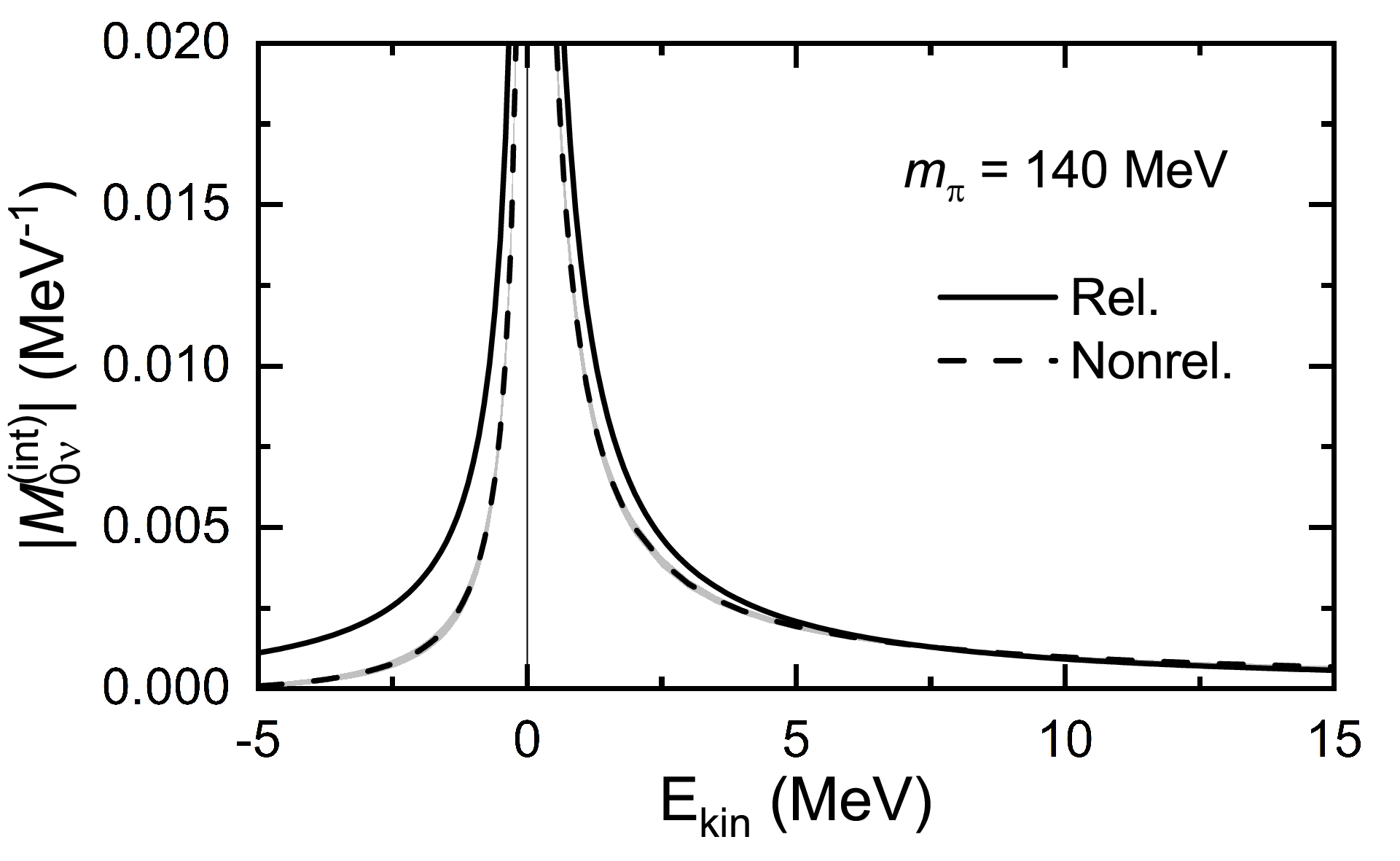}
    \caption{The amplitudes $|\mathcal{M}^{\rm (int)}_{0\nu}|$ at the physical pion mass obtained from the relativistic and nonrelativistic LO pionless EFT, as functions of the center-of-mass kinetic energy $E_{\rm kin}=E-2m_N$ in the initial and final states.
    For the nonrelativistic results, the $nn\rightarrow pp ee$ contact term is fitted to the synthetic datum provided by the generalized Cottingham formula~\cite{Cirigliano2021Phys.Rev.Lett.172002}.
    The effective neutrino mass $m_{\beta\beta}$ is set to 1 MeV.}
    \label{fig3}
\end{figure}
\begin{figure}[!htbp]
    \centering
    \includegraphics[width=0.6\textwidth]{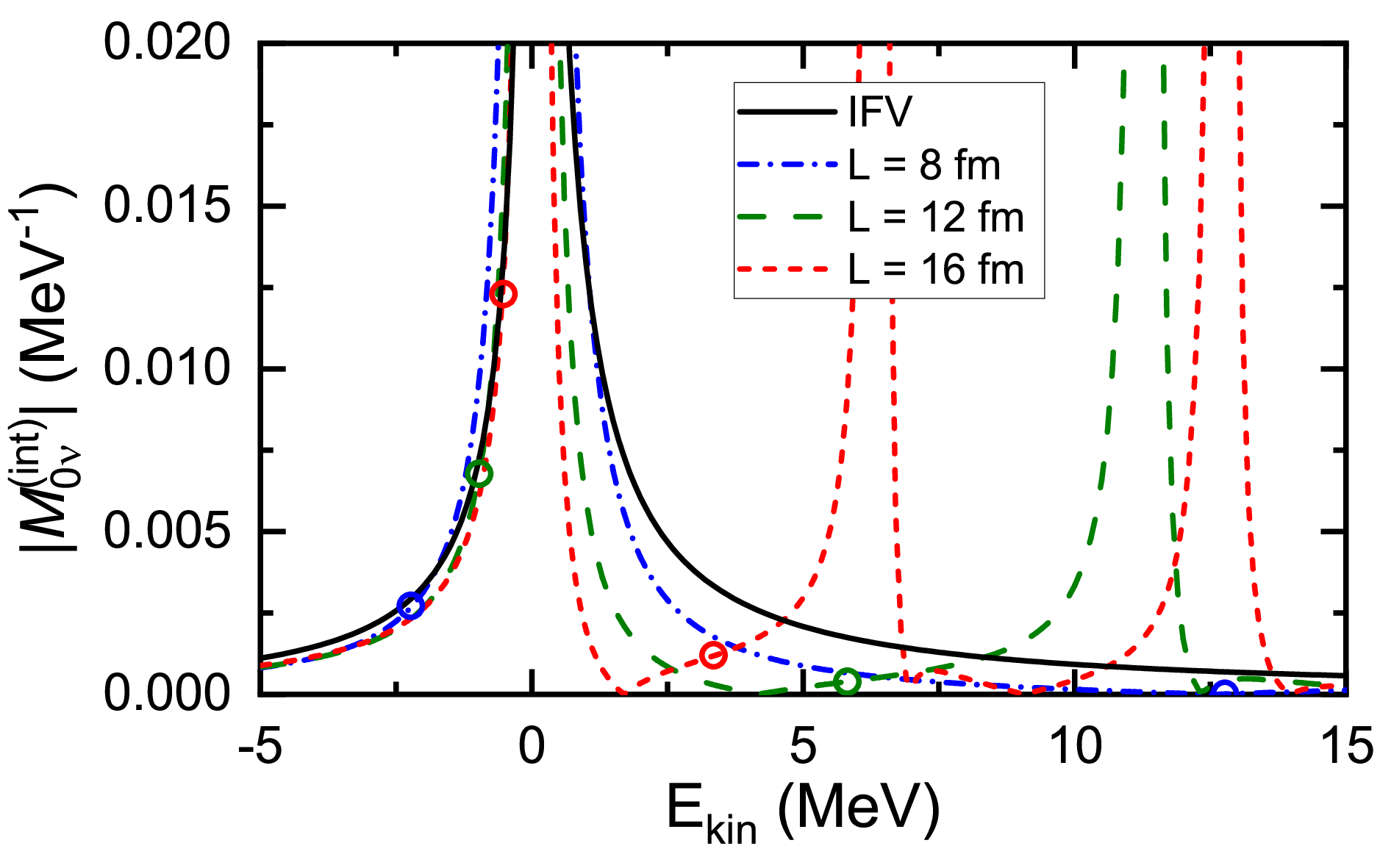}
    \caption{(Color online). The finite-volume quantities $|\mathcal{M}^{({\rm int},L)}_{0\nu}|$ at the physical pion mass obtained from the LO relativistic pionless EFT, as functions of the center-of-mass kinetic energy $E_{\rm kin}=E-2m_N$ in the initial and final states.
    The infinite-volume (IFV) amplitude $|\mathcal{M}^{({\rm int})}_{0\nu}|$ is shown by the solid line.
    The empty circles denote the values of $|\mathcal{M}^{({\rm int},L)}_{0\nu}|$ at the finite-volume energies of the ground states and the first excited states.
    The effective neutrino mass $m_{\beta\beta}$ is set to 1 MeV.}
    \label{fig4}
\end{figure}

We first discuss the $nn\rightarrow ppee$ amplitudes at the physical pion mass.
In Fig.~\ref{fig3}, the absolute value of the infinite-volume amplitude $\mathcal{M}^{(\rm int)}_{0\nu}$ is plotted against the center-of-mass kinetic energy $E_{\rm kin}=E-2m_N$.
The amplitudes obtained from the relativistic formulation are compared to those obtained from the nonrelativistic formulation.
For the latter, dimensional regularization scheme is adopted to regularize the ultraviolet divergence, introducing the renormalization scale $\mu=m_\pi$, and the LEC for the $nn\rightarrow pp ee$ contact term is fitted to the synthetic datum provided by the generalized Cottingham formula~\cite{Cirigliano2021Phys.Rev.Lett.172002}, yielding $\tilde{g}_\nu^{NN}(\mu=m_\pi)=4.09\pm0.21$.
For the energy above the threshold, the nonrelativistic results are consistent with the relativistic ones at $20\%$ level.
For the energy under the threshold, the relative difference between the nonrelativistic and relativistic results grows with decreasing energy.
The amplitude under the threshold is not observable in the continuum, as the kinetic energy cannot be negative.
Nevertheless, it can show up in the matching to the LQCD results, because energies can go below the threshold in finite volumes (e.g., see Fig.~\ref{fig4}).

Figure~\ref{fig4} depicts the volume dependence of the FV quantity $\mathcal{M}^{({\rm int},L)}_{0\nu}$ obtained from the relativistic pionless EFT.
The FV energies of the ground and first-excited states are shown by the empty circles.
The infinite-volume amplitude $\mathcal{M}^{({\rm int})}_{0\nu}$ is also shown for comparison.
For the energy above the threshold $E_{\rm kin}>0$, $\mathcal{M}^{({\rm int},L)}_{0\nu}$ exhibits several singularities in contrast to its infinite-volume counter part $\mathcal{M}^{({\rm int})}_{0\nu}$.
The singularities come from the two-nucleon propagator in Eq.~(\ref{eq.JL}), as its denominator becomes zero for the momentum modes in which two nucleons are on-shell, $E_{\rm kin}=2\sqrt{m_N^2+(2\pi\bm n/L)^2}-2m_N$ with $\bm n\in\mathbb{Z}^3$.
They do not exist in the infinite volume because the on-shell momentum modes contribute to the imaginary part of the propagator instead of being divergent [see Eq.~(\ref{eq.Jinf})].
In between the singularities, the value of $\mathcal{M}^{({\rm int},L)}_{0\nu}$ is generally smaller its infinite-volume counterpart $\mathcal{M}^{({\rm int})}_{0\nu}$.

For the energy below the threshold $E_{\rm kin}<0$, $\mathcal{M}^{({\rm int},L)}_{0\nu}$ behaves smoothly as a function of energy.
In the limit of $L\rightarrow\infty$, the values of $\mathcal{M}^{({\rm int},L)}_{0\nu}$ should approach the infinite-volume amplitude $\mathcal{M}^{({\rm int})}_{0\nu}$.
For the range $L=16$ fm, $\mathcal{M}^{({\rm int},L)}_{0\nu}$ is already close to $\mathcal{M}^{({\rm int})}_{0\nu}$ within $10\%$ at the FV ground-state energies.

\begin{figure}[!htbp]
    \centering
    \includegraphics[width=0.5\textwidth]{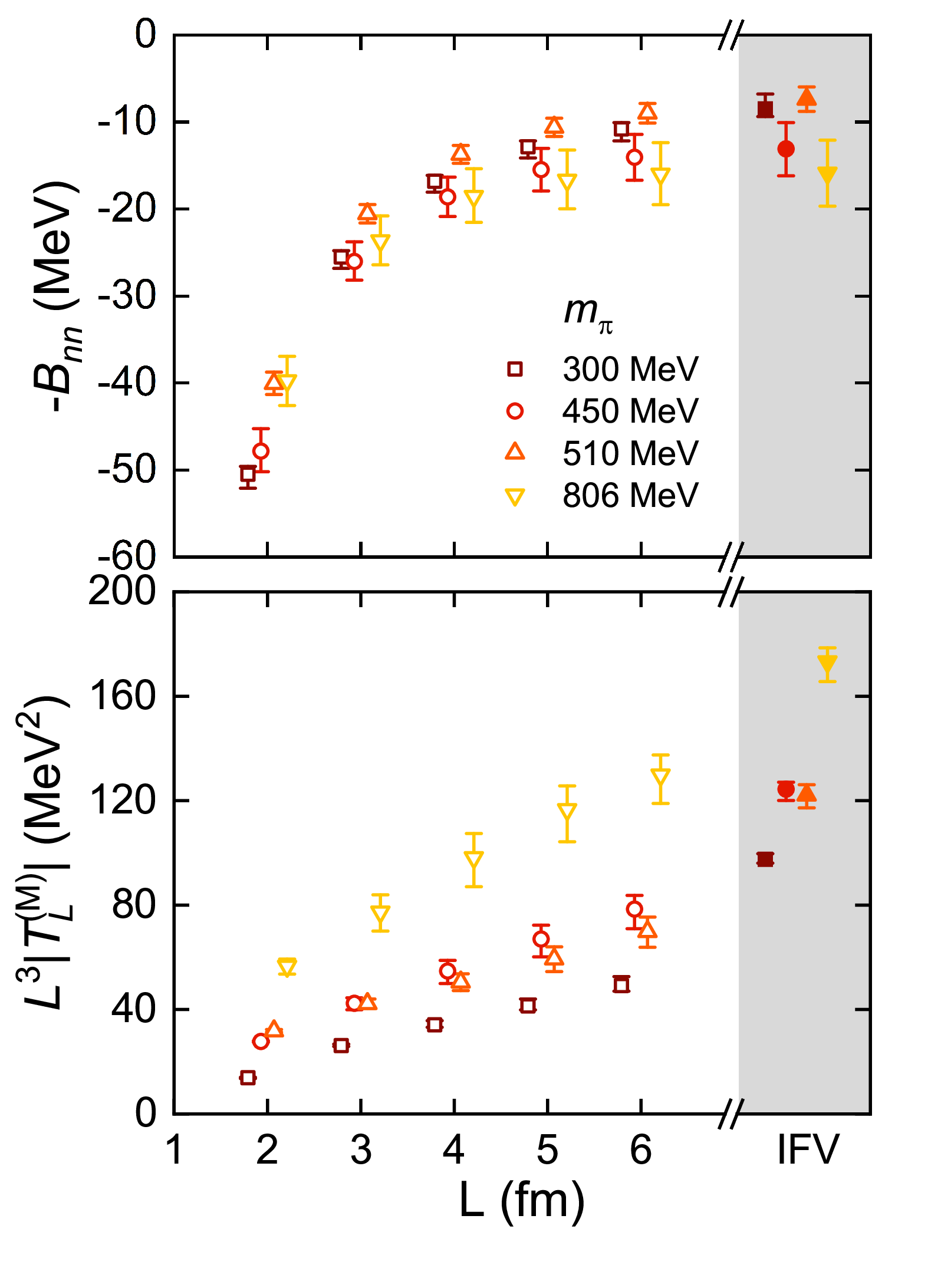}
    \caption{(Color online). The volume dependence of the two-nucleon binding energies $B_{nn}$ and the LO Minkowski matrix elements $\mathcal{T}^{(M)}_L$ for the ground-state-to-ground-state $nn\rightarrow pp ee$ transition at the pion masses $m_\pi=300$, 450, 510, and 806 MeV, obtained from the LO relativistic pionless EFT.
    The results are calculated at integer values of $L$ and slightly shifted in the horizontal direction for clarity.
    Their infinite-volume (IFV) limits are also shown for comparison.
    The error bars are obtained by varying the input data of the two-nucleon binding energies $B_{nn}$ from the LQCD calculations within their margins of errors.
    The effective neutrino mass $m_{\beta\beta}$ is set to 1 MeV.}
    \label{fig5}
\end{figure}

Next, we show the results for the matrix elements at the unphysical pion masses $m_\pi=300$, 450, 510, and 806 MeV.
In these cases, a two-nucleon bound state in the $^1S_0$ channel at each pion mass is predicted by the LQCD calculations~\cite{Yamazaki2015Phys.Rev.D014501,Illa2021Phys.Rev.D054508,Yamazaki2012Phys.Rev.D074514,Beane2013Phys.Rev.D034506,Amarasinghe2023Phys.Rev.D094508}.
Figure~\ref{fig5} depicts the volume dependence of the two-nucleon binding energies $B_{nn}$ and the LO Minkowski matrix elements $\mathcal{T}^{(M)}_L$ for the ground-state-to-ground-state $nn\rightarrow pp ee$ transition, as well as their infinite-volume limits.
The factor $L^3$ is added for $\mathcal{T}^{(M)}_L$ to give the correct normalization in the infinite-volume limit $L\rightarrow\infty$.
The Minkowski matrix element generally decreases when the pion mass becomes smaller.

For each pion mass, the binding energy $B_{nn}$ becomes significantly large for small box sizes $L\lesssim3$ fm and comes close to the infinite-volume value at $L=6$ fm.
However, this is not the case for the $nn\rightarrow ppee$ matrix element.
At the heaviest pion mass $m_\pi=806$ MeV, it increases from 33\% of the infinite-volume limit at $L=2$ fm to 75\% at $L=6$ fm.
At the lightest pion mass $m_\pi=300$ MeV, it increases from 14\% of the infinite-volume limit at $L=2$ fm to only 50\% at $L=6$ fm.
The value of $L^3\mathcal{T}^{(M)}_L$ increases slowly with increasing box size, so a much larger box size is needed to approach the infinite-volume limit.

The different volume dependence between the binding energy and the $nn\rightarrow ppee$ matrix element is due to the fact that the strong interaction is short-range while the neutrino exchange is long-range.
The photon exchange responsible for the electromagnetic interactions is also long-range, and it is known that the FV corrections for the electromagnetic interactions exhibit a power-law scaling with volume~\cite{Davoudi2014Phys.Rev.D054503}, instead of an exponential scaling for the short-range strong interactions.
Besides approaching the infinite-volume limit by increasing the box size, one could also extrapolate the results obtained using relatively small box sizes.
We extrapolate the values of $L^3\mathcal{T}^{(M)}_L$ at $L=4$, 5, 6 fm to infinite volume by considering the leading $O(1/L)$ correction.
The extrapolation reduces the difference against the infinite-volume limit, but systematic deviation remains.
At the heaviest pion mass $m_\pi=806$ MeV, the extrapolated result overestimates the infinite-volume value by about 10\%.
While at the lightest pion mass $m_\pi=300$ MeV, the extrapolated result underestimates the infinite-volume value by about 20\%.

\begin{figure}[!htbp]
    \centering
    \includegraphics[width=0.5\textwidth]{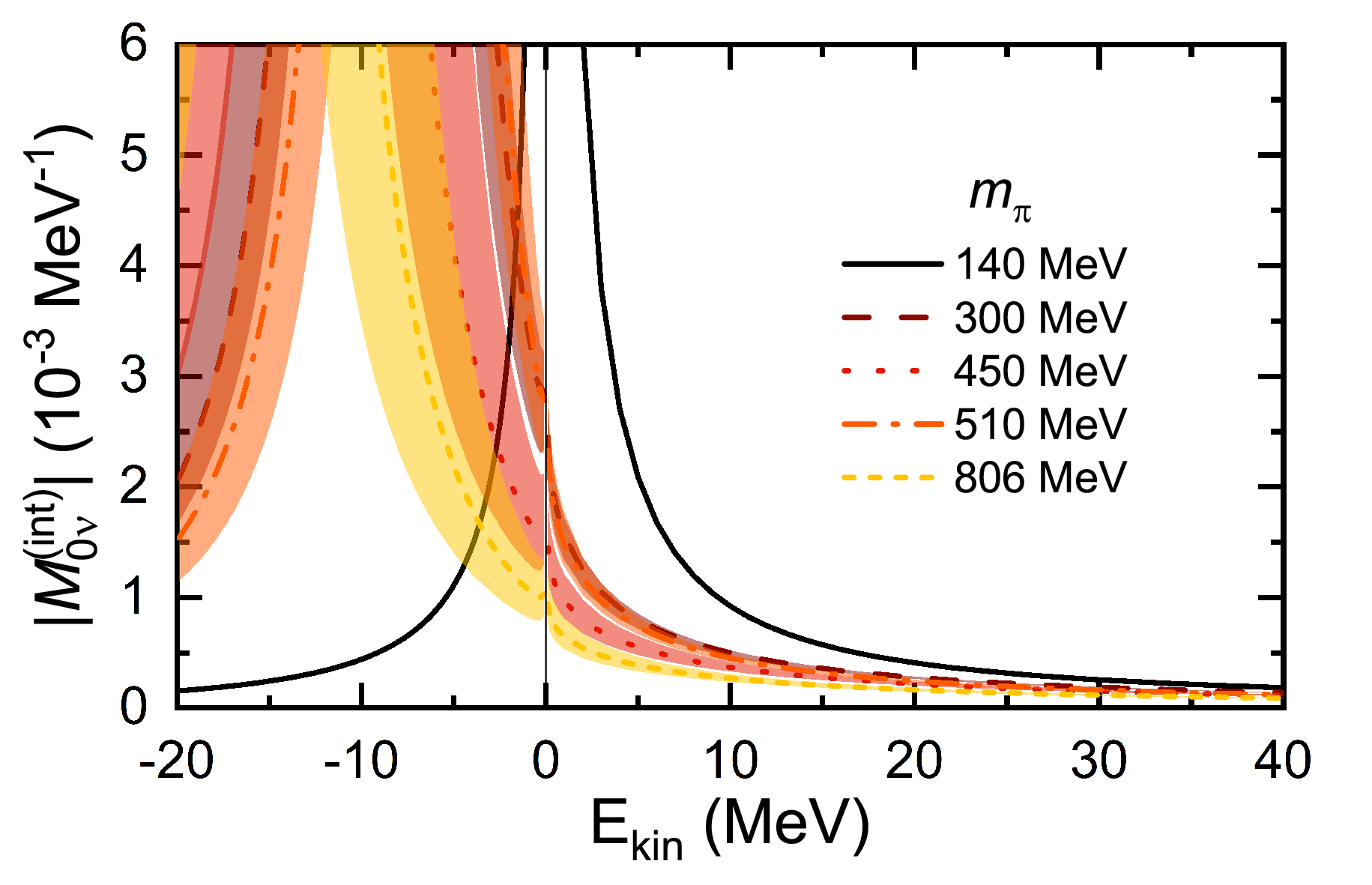}
    \caption{(Color online). The amplitudes $|\mathcal{M}^{({\rm int})}_{0\nu}|$ at different pion masses obtained from the LO relativistic pionless EFT, as functions of the center-of-mass kinetic energy $E_{\rm kin}=E-2m_N$ in the initial and final states.
    The shaded uncertainties are obtained by varying the input data of the two-nucleon binding energies $B_{nn}$ from the LQCD calculations within their margins of errors.
    The effective neutrino mass $m_{\beta\beta}$ is set to 1 MeV.
}
    \label{fig6}
\end{figure}
Figure~\ref{fig6} depicts the absolute value of the amplitude $\mathcal{M}^{({\rm int})}_{0\nu}$ at different pion masses.
The shaded uncertainties are obtained by varying the input data from the LQCD calculations within their margins of errors.
For the energy above the threshold, the amplitudes $|\mathcal{M}^{({\rm int})}_{0\nu}|$ at the unphysical pion masses are significantly smaller than that at the physical pion mass.
$|\mathcal{M}^{({\rm int})}_{0\nu}|$ drops rapidly with increasing energy, which is similar at both the physical and unphysical masses.

For the energy below the threshold, however, the amplitude $|\mathcal{M}^{({\rm int})}_{0\nu}|$ exhibits very different behavior at the unphysical pion masses compared to that at the physical pion mass.
In particular, the amplitude $|\mathcal{M}^{({\rm int})}_{0\nu}|$ diverges at the energy of two-nucleon bound state at each unphysical pion mass, because the bound-state energy is the pole of the two-nucleon scattering amplitude $M_S(E)$.
For the $nn\rightarrow pp ee$ transition between bound states, the scattering amplitude $|\mathcal{M}^{({\rm int})}_{0\nu}|$ (and also $|\mathcal{M}^{({\rm int}, L)}_{0\nu}|$) is not well-defined and one should directly calculate the Minkowski matrix element $\mathcal{T}^{(M)}$ ($\mathcal{T}^{(M)}_L$) using bound-state wave functions.
Such divergence does not exist at the physical pion mass since there is no two-nucleon bound state in the $^1S_0$ channel.

\begin{figure*}[!htbp]
    \centering
    \includegraphics[width=0.9\textwidth]{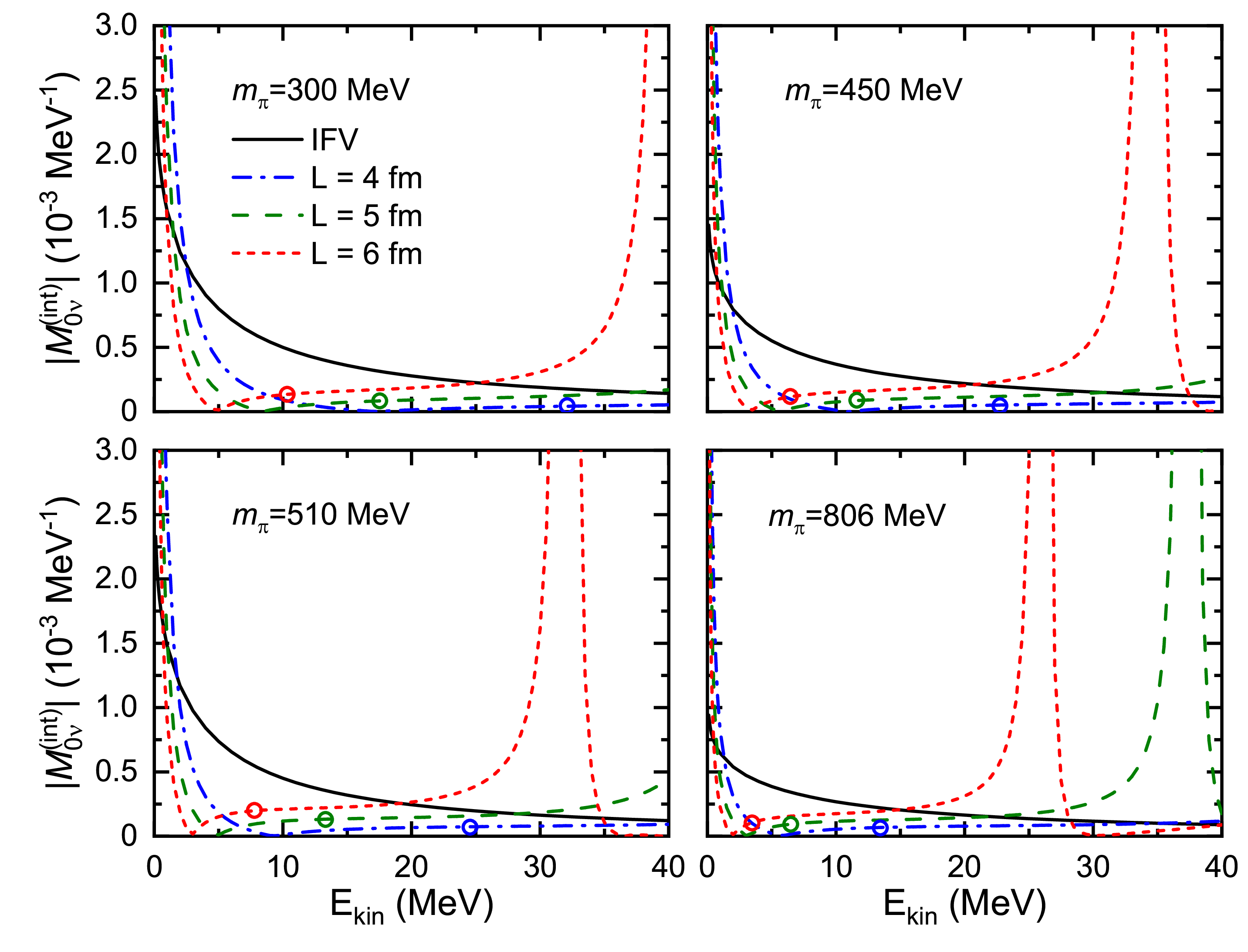}
    \caption{(Color online). The finite-volume quantities $|\mathcal{M}^{({\rm int},L)}_{0\nu}|$ at the unphysical pion masses obtained from the LO relativistic pionless EFT, as functions of the center-of-mass kinetic energy $E_{\rm kin}=E-2m_N$ in the initial and final states.
    The empty circles denote the values of $|\mathcal{M}^{({\rm int},L)}_{0\nu}|$ at the finite-volume energies of the ground states and the first excited states.
    The infinite-volume (IFV) amplitudes $|\mathcal{M}^{({\rm int})}_{0\nu}|$ are also shown by the solid lines.
    Here, the results are obtained using the central values of the two-nucleon binding energies $B_{nn}$ from the LQCD calculations as inputs.
    The effective neutrino mass $m_{\beta\beta}$ is set to 1 MeV.}
    \label{fig7}
\end{figure*}
In Fig.~\ref{fig7}, the FV quantities $|\mathcal{M}^{({\rm int},L)}_{0\nu}|$ at the unphysical pion masses with the box sizes $L=4,5,6$ fm are depicted in comparison with the results in the infinite volume, as functions of the center-of-mass kinetic energy above the threshold.
Here, the results are obtained using the central values of LEC in Table~\ref{tab1} and the uncertainties from the LQCD inputs are not shown.
The behavior of $|\mathcal{M}^{({\rm int},L)}_{0\nu}|$ at the unphysical pion masses are similar to that at the physical pion mass except for the locations of singularities. 
This is because $|\mathcal{M}^{({\rm int},L)}_{0\nu}|$ diverges at the neighborhood of the integer multiples of $4\pi^2/(L^2m_N)$, at which the two nucleons can become on-shell in the FV two-loop integral $J^L$ [Eq.~(\ref{eq.JL})].
As a result, the singularities are denser at heavier pion mass, since the nucleon mass increases with increasing pion mass, and for larger spatial volumes.
The FV energies of the ground states and the first excited states in the different spatial volumes are shown by the empty circles. 
The values of $|\mathcal{M}^{({\rm int},L)}_{0\nu}|$ at the FV energies are significantly smaller than the infinite-volume amplitude $|\mathcal{M}^{({\rm int})}_{0\nu}|$.

\begin{table*}
	\centering
	\caption{The Minkowski matrix elements $\mathcal{T}^{(M)}_L$ for the ground-state-to-ground-state and first-excited-state-to-first-excited-state $nn\rightarrow pp ee$ transitions at the pion masses $m_\pi=300$, 450, 510, and 806 MeV, predicted by the LO relativistic pionless EFT.
    The finite-volume energies $E_0$ and $E_1$ for the ground states and the first excited states are also shown, respectively.
    The uncertainties are obtained by varying the input data of the two-nucleon binding energies $B_{nn}$ from the LQCD calculations within their margins of errors.}
	\label{tab2}
	\begin{tabular}{cccccc}
	    \hline\hline
		$m_\pi$ (MeV) & $L$ (fm) & $E_0$ (MeV) & $\mathcal{T}^{(M)}_L(E_0,E_0)$ (MeV$^{5}$) & $E_1$ (MeV) & $\mathcal{T}^{(M)}_L(E_1,E_1)$ (MeV$^5$)\\ \hline
		    & 4 & -16.8$\left(^{+1.3}_{-0.7}\right)$ & 4.1$\left(^{+0.2}_{-0.1}\right)\times10^6$ & 31.6$\left(^{+0.9}_{-1.5}\right)$ & 1.9$\left(^{+0.1}_{-0.2}\right)\times10^8$\\
		300 & 5 & -12.9$\left(^{+1.3}_{-0.7}\right)$ & 2.5$\left(^{+0.2}_{-0.1}\right)\times10^6$ & 17.3$\left(^{+0.7}_{-1.2}\right)$ & 1.8$\left(^{+0.1}_{-0.2}\right)\times10^8$\\
		    & 6 & -10.8$\left(^{+1.4}_{-0.7}\right)$ & 1.7$\left(^{+0.1}_{-0.1}\right)\times10^6$ & 10.2$\left(^{+0.5}_{-0.9}\right)$ & 1.6$\left(^{+0.2}_{-0.3}\right)\times10^8$\\\hline
		    
		    & 4 & -18.6$\left(^{+2.3}_{-2.3}\right)$ & 6.6$\left(^{+0.5}_{-0.6}\right)\times10^6$ & 23.7$\left(^{+2.2}_{-1.7}\right)$ & 2.3$\left(^{+0.4}_{-0.3}\right)\times10^8$\\
		450 & 5 & -15.5$\left(^{+2.5}_{-2.4}\right)$ & 4.1$\left(^{+0.3}_{-0.4}\right)\times10^6$ & 11.4$\left(^{+1.5}_{-1.2}\right)$ & 1.7$\left(^{+0.4}_{-0.4}\right)\times10^8$\\
		    & 6 & -14.1$\left(^{+2.6}_{-2.6}\right)$ & 2.8$\left(^{+0.2}_{-0.3}\right)\times10^6$ & 6.3$\left(^{+1.1}_{-0.7}\right)$  & 1.2$\left(^{+0.5}_{-0.3}\right)\times10^8$\\\hline
		    
		    & 4 & -13.7$\left(^{+1.0}_{-1.0}\right)$ & 6.1$\left(^{+0.4}_{-0.4}\right)\times10^6$ & 24.2$\left(^{+1.3}_{-1.2}\right)$ & 3.0$\left(^{+0.2}_{-0.2}\right)\times10^8$\\
		510 & 5 & -10.6$\left(^{+1.1}_{-1.0}\right)$ & 3.7$\left(^{+0.3}_{-0.3}\right)\times10^6$ & 13.1$\left(^{+1.0}_{-0.9}\right)$ & 2.5$\left(^{+0.3}_{-0.3}\right)\times10^8$\\
		    & 6 & -9.0$\left(^{+1.2}_{-1.1}\right)$  & 2.5$\left(^{+0.2}_{-0.2}\right)\times10^6$ & 7.7$\left(^{+0.8}_{-0.7}\right)$  & 2.0$\left(^{+0.4}_{-0.3}\right)\times10^8$\\\hline
		    
		    & 4 & -18.6$\left(^{+3.0}_{-3.2}\right)$ & 1.2$\left(^{+0.1}_{-0.1}\right)\times10^7$ & 13.1$\left(^{+1.8}_{-1.4}\right)$ & 2.6$\left(^{+0.6}_{-0.5}\right)\times10^8$\\
		806 & 5 & -16.7$\left(^{+3.3}_{-3.5}\right)$ & 7.2$\left(^{+0.6}_{-0.8}\right)\times10^6$ & 6.3$\left(^{+1.1}_{-0.8}\right)$  & 1.6$\left(^{+0.7}_{-0.5}\right)\times10^8$\\
		    & 6 & -16.0$\left(^{+3.5}_{-3.6}\right)$ & 4.6$\left(^{+0.3}_{-0.4}\right)\times10^6$ & 3.4$\left(^{+0.7}_{-0.5}\right)$  & 9.3$\left(^{+0.6}_{-0.4}\right)\times10^8$\\
		\hline\hline		
	\end{tabular}
\end{table*}
In Table~\ref{tab2},  we provide the values of the LO Minkowski matrix elements $\mathcal{T}^{(M)}_L$ for the ground-state-to-ground-state and first-excited-state-to-first-excited-state $nn\rightarrow pp ee$ transitions in finite volumes with $L=4$, 5, and 6 fm.
The precision of the predicted matrix elements is mostly within 10\%-20\%.
This precision is in accordance with the precision of the two-nucleon binding energies $B_{nn}$ from the LQCD calculations (Table~\ref{tab1}) used as inputs.
In addition to the uncertainty from the LQCD input, there are also uncertainties from the truncation of EFT and the estimation of the LO LEC $\tilde{g}_\nu^{NN}$, as discussed in Sec.~\ref{sec.IIA}.
For the former, the uncertainty arises from neglecting the strong potential and neutrino potential beyond LO, expected to be $O(Q/m_\pi)$.
For the latter, it takes into account the known LO contribution from the coupling of pions to axial currents, and its uncertainty comes from the possible unknown short-range contributions.
This unknown short-range contributions is expected to be subleading based on the comparison between the pionless and chiral EFTs.
Therefore, we expect the truncation uncertainty of the predictions of relativistic pionless EFT is of the order of $O(Q/m_\pi)$, with $Q$ estimated by the two-nucleon binding energy $\sqrt{m_N B_{nn}}$ or the inverse scattering length $a^{-1}$.
For $m_\pi=300$, 450, 510, and 806 MeV, the truncation uncertainties are expected to be of the order of $32\%$, $28\%$, $19\%$, and $20\%$, respectively.
In general, the truncation uncertainty should become smaller for heavier pion mass for the relativistic pionless EFT.

Finally, we present a comparison with the first evaluation of the ground-state-to-ground-state $nn\rightarrow pp ee$ matrix element on the lattice with $L=4.6$ fm at $m_\pi=806$ MeV, achieved by NPLQCD Collaboration~\cite{Davoudi2024Phys.Rev.D114514}.
For the $^1S_0$ two-nucleon energy at this pion mass, there exists a discrepancy between the older results~\cite{Beane2013Phys.Rev.D034506,Berkowitz2017Phys.Lett.B285,Wagman2017Phys.Rev.D114510} and the latest result~\cite{Amarasinghe2023Phys.Rev.D094508} by NPLQCD Collaboration.
Such discrepancy is suspected to be due to the misidentification of the two-nucleon spectrum through ``false plateaus" in the older works, yielding a deeply bound two-nucleon state~\cite{Drischler2021Prog.Part.Nucl.Phys.103888,Davoudi2021Phys.Rep.174,Tews2022FewBodySystems67}.
Several newer works~\cite{Francis2019Phys.Rev.D074505,Horz2021Phys.Rev.C014003,Amarasinghe2023Phys.Rev.D094508,Detmold2024arXiv} have not identified such deeply bound two-nucleon state.
Nevertheless, there are several explanations and this issue is still not completely settled~\cite{Drischler2021Prog.Part.Nucl.Phys.103888,Davoudi2021Phys.Rep.174,Tews2022FewBodySystems67}.
Here, we used both the older and the latest results for the two-nucleon energy from Refs.~\cite{Beane2013Phys.Rev.D034506,Amarasinghe2023Phys.Rev.D094508} as inputs of the EFT (Table \ref{tab1}), as they are both consistent with the one yielded in the $nn\rightarrow pp ee$ calculation~\cite{Davoudi2024Phys.Rev.D114514}.
The results are shown below,
\begin{equation}
    \begin{split}
    &\left|\mathcal{T}^{(M)}_{L}\right|_{\rm EFT}=\left\{
    \begin{split}
    &8.7\left(^{+0.7}_{-1.0}\right)\times 10^6\ {\rm MeV}^5\quad(B_{nn}\simeq17\ {\rm MeV})\\
    &1.7\left(^{+0.7}_{-0.4}\right)\times 10^6\ {\rm MeV}^5\quad(B_{nn}\simeq3\ {\rm MeV})\\
    \end{split},
    \right.\\
    &\left|\mathcal{T}^{(M)}_{L}\right|_{\rm LQCD}=1.75\left(^{+0.36}_{-0.36}\right)\times 10^6\ {\rm MeV}^5.
    \end{split}
\end{equation}
If the latest results of the two-nucleon energy~\cite{Amarasinghe2023Phys.Rev.D094508} is adopted, the EFT prediction of the matrix element is consistent with the LQCD result, within the uncertainty coming from the inputs.
However, if the deeply bound two-nucleon energy from the older calculation~\cite{Beane2013Phys.Rev.D034506} is adopted, the matrix element is significantly larger than the LQCD result.
This is because the neutrino exchange potential behaves as $1/r$ in the coordinate space, and the larger the binding energy, the more compact the two-nucleon system.
In addition, the physical value of the axial coupling constant $g_A=1.27$ is used here, while $g_A$ should slightly decrease with increasing pion mass~\cite{Chang2018Nature91} and this could slightly decrease the present EFT prediction.
Nevertheless, the agreement between the present EFT prediction (using the newest results of the two-nucleon energy~\cite{Amarasinghe2023Phys.Rev.D094508}) and the first LQCD evaluation for the $nn\rightarrow pp ee$ matrix element is very encouraging.
To be more conclusive, future benchmarks should be carried out after the LQCD calculations reduce the uncertainties in the two-nucleon energy.
In addition, we anticipate more LQCD calculations of the $nn\rightarrow pp ee$ matrix elements at different pion masses or finite volumes.
Then, the systematic comparison between the EFT matrix elements and the LQCD ones could be a stringent benchmark for the validity of EFT predictions on the $nn\rightarrow ppee$ process.

%*********************************************************%
%------------------------Summary--------------------------%
%*********************************************************%
\section{Summary}
In this work, the neutrinoless double-beta decay process $nn\rightarrow ppee$ within the light Majorana-neutrino exchange scenario is studied in a finite volume based on the leading-order relativistic pionless EFT.
The finite-volume Minkowski matrix elements of the $nn\rightarrow pp ee$ process are predicted for the pion masses $m_\pi=300$, 450, 510, and 806 MeV, at which the LQCD calculations of the two-nucleon energies exist.
These results can be directly compared to the results from LQCD calculations of the $nn\rightarrow pp ee$ process at the same pion masses.

The previous studies~\cite{Davoudi2021Phys.Rev.Lett.152003, Davoudi2022Phys.Rev.D094502} presented the matching framework between the finite-volume matrix elements from LQCD and the infinite-volume scattering amplitude from the nonrelativistic pionless EFT for the $nn\rightarrow pp ee$ process.
The scattering amplitudes and finite-volume Minkowski matrix elements of the $nn\rightarrow pp ee$ process are calculated at the physical pion masses~\cite{Davoudi2022Phys.Rev.D094502}, where the size of the LO $nn\rightarrow pp ee$ contact term is determined by the generalized Cottingham formula~\cite{Cirigliano2021Phys.Rev.Lett.172002}.
However, such determination of the contact term is not applicable at the unphysical pion masses.
Different from the nonrelativistic studies, the present work presents a relativistic study, where the size of the contact term is determined by integrating out the pion contributions to the long-range neutrino potential in the relativistic chiral EFT.
This is possible thanks to the fact that the long-range $nn\rightarrow pp ee$ amplitudes are renormalizable at leading order in the relativistic chiral EFT~\cite{Yang2024Phys.Lett.B138782}, in contrast to the nonrelativistic case.
The obtained amplitudes at the physical pion mass are consistent with the previous nonrelativistic results at 20\% level.
In addition, the $nn\rightarrow ppee$ processes at the unphysical pion masses $m_\pi=300$, 450, 510, and 806 MeV are studied in a finite volume for the first time, based on the relativistic pionless EFT, using the two-nucleon energies from the existing LQCD calculations~\cite{Yamazaki2015Phys.Rev.D014501,Illa2021Phys.Rev.D054508,Yamazaki2012Phys.Rev.D074514,Beane2013Phys.Rev.D034506,Amarasinghe2023Phys.Rev.D094508}.

At the unphysical pion masses, the renormalization-group invariance of the leading-order Minkowski matrix elements is confirmed.
Then, the matrix elements are predicted in several different volumes to investigate their volume dependence.
It is found that a much larger volume than those implemented in the present LQCD studies of two-nucleon systems (typically with cubic-box sizes in the range of $L=$4-6 fm) is required to approach the infinite-volume limits of the $nn\rightarrow ppee$ matrix elements, due to the long-range nature of neutrino exchange.
The finite-volume results can be improved by the extrapolation considering the leading $O(1/L)$ correction, but systematic deviations from the infinite-volume limit remain for about 10\%-20\%, depending on the pion mass.

Finally, the relativistic pionless EFT predictions of the Minkowski matrix elements in several finite volumes are presented for the ground-state-to-ground-state and first-excited-state-to-first-excited-state $nn\rightarrow ppee$ transitions at the pion masses $m_\pi=300$, 450, 510, and 806 MeV.
These results allow direct benchmarks between EFT and LQCD on the $nn\rightarrow ppee$ process, especially at the heavy pion masses that are numerically more favorable for LQCD.
In particular, the EFT predictions for $m_\pi=806$ MeV are compared with the first LQCD evaluation of the ground-state-to-ground-state $nn\rightarrow ppee$ matrix element at a finite volume of $L=4.6$ fm~\cite{Davoudi2024Phys.Rev.D114514}.
Using the latest LQCD value of two-nucleon energy in a same lattice setup ~\cite{Amarasinghe2023Phys.Rev.D094508} as inputs, the relativistic pionless EFT yields a $nn\rightarrow ppee$ matrix element in good agreement with the LQCD evaluation.
This is not the case if the deeply bound two-nucleon energy from the older LQCD calculation~\cite{Beane2013Phys.Rev.D034506} is used.

The present results motivate future studies of the $nn\rightarrow ppee$ process from LQCD at different pion masses and finite volumes.
In addition, the present leading-order study on the $nn\rightarrow ppee$ process in a finite volume also provides the basis for such studies at higher orders, where the LECs associated with subleading lepton-number-breaking contact terms have to be determined via matching to LQCD calculations.

% Acknowledgement
\begin{acknowledgments}
This work has been supported in part by the National Natural Science Foundation of China (Grants No. 123B2080, No. 12141501, No. 11935003), and the High-performance Computing Platform of Peking University.
We acknowledge the funding support from the State Key Laboratory of Nuclear Physics and Technology, Peking University (Grant No. NPT2023ZX03).
\end{acknowledgments}

%\bibliography{ref0nbbFV}


\end{document}